\documentclass[11pt, a4paper]{article}

\usepackage[T1]{fontenc}
\usepackage[utf8]{inputenc} 
\usepackage{graphicx,color} 
\usepackage{epsfig} 
\usepackage{times} 
\usepackage{amsmath}
\usepackage{amssymb}
\usepackage{caption}
\usepackage{subcaption}
\usepackage{stackrel}

\usepackage{url}
\usepackage{multirow}
\usepackage{cite}

\usepackage{algorithm}
\usepackage{pgf}
\usepackage{pgfplots}
\usepgfplotslibrary{fillbetween}
\usepackage{tikz}
\usetikzlibrary{arrows,automata}
\usetikzlibrary{intersections}
\usetikzlibrary{calc}
\usetikzlibrary{shapes}
\usetikzlibrary{positioning}
\usetikzlibrary{decorations.text}
\pgfdeclarelayer{background}
\pgfdeclarelayer{foreground}
\pgfsetlayers{background,main,foreground}
\usetikzlibrary{calc,patterns,decorations.pathmorphing,decorations.markings}

\usepackage{enumitem}
\renewcommand{\theenumi}{\arabic{enumi}}

\newcommand{\dumin}{\Delta u_{\rm min}}
\newcommand{\dumax}{\Delta u_{\rm max}}
\newcommand{\umin}{u_{\rm min}}
\newcommand{\umax}{u_{\rm max}}
\newcommand{\ymin}{y_{\rm min}}
\newcommand{\ymax}{y_{\rm max}}
\newcommand{\zmin}{z_{\rm min}}
\newcommand{\zmax}{z_{\rm max}}
\newcommand{\eps}{\zeta}
\newcommand{\NN}{{\mathcal N}}

\newcommand{\XX}{{\mathcal X}}
\newcommand{\WW}{{\mathcal W}}
\newcommand{\VV}{{\mathcal V}}
\newcommand{\FF}{{\mathcal F}}
\newcommand{\EE}{{\mathcal E}}

\newcommand{\GG}{{\mathcal G}}
\newcommand{\Ss}{{\mathcal S}}

\newtheorem{proposition}{Proposition}
\newtheorem{example}{Example}

\newcommand{\rr}{\mathbb{R}}
\newcommand{\st}{\mbox{s.t.~} }
\newcommand{\eqdef}{\triangleq}
\newcommand{\proof}{\noindent{\em Proof}.~}
\newcommand{\QED}{$\Box$}
\newcommand{\cvd}{\hfill\QED}

\newcommand{\beqar}{\begin{eqnarray}}
\newcommand{\eeqar}{\end{eqnarray}}
\newcommand{\beqarno}{\begin{eqnarray*}}
\newcommand{\eeqarno}{\end{eqnarray*}}
\newcommand{\ba}[1]{\begin{array}{#1}}
\newcommand{\ea}{\end{array}}

\newcommand{\sign}{\mathop{\rm sign}\nolimits}
\newcommand{\col}{\mathop{\rm col}\nolimits}
\newcommand{\sat}{\mathop{\rm sat}\nolimits}
\newcommand{\softsat}{\sigma}

\newcommand{\maxzero}[1]{[#1]_+}

\begin{document}
    \title{Worst-case Nonlinear Regression with Error Bounds}
    
	\author{Alberto Bemporad 
    \thanks{The author is with the IMT School for Advanced Studies, Piazza San Francesco 19, Lucca, Italy. Email: \texttt{\scriptsize alberto.bemporad@imtlucca.it}.
    This work was funded by the European Union (ERC Advanced Research Grant COMPACT, No. 101141351). Views and opinions expressed are however those of the authors only and do not necessarily reflect those of the European Union or the European Research Council. Neither the European Union nor the granting authority can be held responsible for them.
}}

\maketitle
\thispagestyle{empty}
	
\begin{abstract}
We propose an active-learning method for nonlinear minimax regression. Given a nonlinear function that can be arbitrarily evaluated over a compact set, we fit a surrogate model, such as a feedforward neural network, by minimizing the maximum absolute approximation error. To handle the nonsmoothness of this worst-case loss, we introduce a smooth $L_\infty$ approximation that enables efficient gradient-based training. The training set is iteratively enriched by querying points of largest error via global optimization. We also derive constant and input-dependent worst-case error bounds over the entire input domain. The approach is validated on approximations of nonlinear functions and nonconvex sets, uncertain models of nonlinear dynamics, and explicit model predictive control laws. A Python library is available at \url{https://github.com/bemporad/maxfit}.
\end{abstract}

\textbf{Keywords}:\\
Worst-case nonlinear regression, $L_\infty$-regression, minimax regression, neural networks, uncertainty bounds, approximate explicit model predictive control.

\section{Introduction}
Function approximation methods are fundamental in control systems engineering, both to obtain control-oriented models of dynamical systems and to simplify complex nonlinear control laws for online implementation. They can be used either to learn models or control laws directly from data, or to simplify a known model or control function. In both cases, the typical approach is to collect a sufficiently large set of samples and solve a nonlinear regression problem, for example by training a feedforward neural network (NN) by minimizing the mean squared error (MSE), possibly under regularization on the model parameters.

In control applications, however, it is often crucial to quantify a bound on the approximation error between the true function, when known, and the learned model. 
Such bounds are needed, for example, in robust control design to characterize modeling errors treated as additive disturbances, and when approximating control laws to guarantee closed-loop stability and constraint satisfaction despite approximation errors. This issue is particularly relevant for approximating model predictive control (MPC) laws to reduce online computational effort, typically using NNs~\cite{PZ95,CSALKP18,
KL20}. However, most existing methods do not guarantee bounds on the worst-case approximation error (WCE) between the exact MPC law and its approximation. The recent contribution~\cite{ATSMF24} addresses this problem by exploiting Lipschitz continuity, assuming the availability of a sufficiently large training dataset.

In system identification for robust control, set-membership methods 
have been used to obtain models with guaranteed bounded error predictions, including nonlinear models~\cite{MN04}, often under assumptions on the system dynamics. Similar ideas have been applied to approximate nonlinear MPC laws in~\cite{CFM08} under Lipschitz continuity assumptions. Gaussian processes~\cite{RW06} offer an alternative by providing models with quantified uncertainty, but they rely on restrictive model classes and yield only probabilistic guarantees.

Robustness in nonlinear regression has been studied in the statistics literature mainly to mitigate the effect of outliers in training data~\cite{HR09}. Here, instead, the goal is to minimize and quantify the WCE over {\it all} possible inputs in a given set. This problem is related to uncertainty quantification and conformal prediction~\cite{AB23}, which typically provide probabilistic confidence intervals rather than worst-case guarantees~\cite{BCRT21}. Achieving the latter requires different techniques; for example, in~\cite{FMP22} semidefinite programming is used to certify the robustness of NN predictions against input uncertainties.

The standard workflow to obtain nonlinear models with bounded prediction errors is to first identify a model and then estimate its approximation error. However, a small {\it average} error, such as a low MSE, does not necessarily imply a small {\it worst-case} error. This motivates methods that directly minimize the WCE during training and reliably quantify it afterward.

{\it Minimax regression}, also known as {\it Chebyshev regression} or {\it $L_\infty$ estimation}~\cite[Ch.~7]{Pow81}, addresses this need by minimizing the maximum deviation between the true function and its approximation over a dataset or a domain. Despite its potential for robust system identification and control, it has received limited attention in the machine learning and control communities, with most existing results restricted to linear models. This is partly due to the nonsmooth optimization problems it entails and to the strong dependence of the solution quality on the choice of the training samples.

\subsection{Contribution}
In this paper, we consider fitting a parametric model, such as a NN, to a known nonlinear function using an $\ell_\infty$ loss over a finite set of samples, with the goal of minimizing the WCE over a compact input set. To enable the use of efficient quasi-Newton methods such as L-BFGS~\cite{LN89} together with automatic differentiation, we replace the nonsmooth maximum operator with a smooth (but arbitrarily accurate) approximation and solve the resulting problem. We reduce the dependence of the WCE on training samples by proposing an active-learning strategy that iteratively adds points of maximum approximation error, identified via global optimization.

A key difference with respect to active-learning methods for nonlinear regression and classification, which rely on empirical acquisition functions to select potentially informative data~\cite{Set12,Bem23c}, is that here the true function can be evaluated at arbitrary inputs during the acquisition process to explicitly identify a point of maximum error. The approach is therefore closely related to corner-case detection via numerical optimization~\cite{MFGB23}.

In addition, we propose techniques to quantify the WCE over the entire input set after training, in the form of either constant or input-dependent upper and lower bounds on the approximation error. Finally, we show how the proposed approach can be used to learn sets (e.g., minimally conservative convex inner approximations of given nonconvex sets), uncertain (generally nonlinear) models of nonlinear dynamical systems (e.g., derived from physical principles), and approximate explicit MPC laws with guaranteed bounds on the control error introduced by the approximation.

\subsection{Notation}
We denote by $\maxzero{x}=\max(x,0)$ the positive part of $x\in\rr$ and by $\max(v)=\max_{i=1,\ldots,n} (v_i)$ the maximum component of a vector $v\in\rr^n$. Given vectors $v_i\in\rr^{n_i}$, $i=1,\ldots,N$, $n_i\geq 1$, we denote by $v=\col(v_1,\ldots,v_N)\in\rr^n$ their vertical concatenation. Throughout the paper, with a slight abuse of notation,
we define the sign function as $\sign(x)=-1$ if $x\leq 0$ and $\sign(x)=1$ if $x>0$. 

\section{Minimax Nonlinear Regression}
\label{sec:worst-case-regression}
Given a nonlinear function $f:\XX\to\rr$ over a compact set $\XX\subset\rr^n$ and a class of functions $\FF$,
our goal is to find a simpler function $\hat f:\XX\to\rr$, $\hat f\in\FF$, that minimizes the WCE between $f(x)$ and $\hat f(x)$ over $\XX$, and to quantify such an error. 

Let $\theta$ be the vector of parameters identifying the function $\hat f$ within $\FF$, 
$\theta\in\rr^{n_\theta}$. For example, $\theta$ can collect the weights and bias terms of a NN, the coefficients of a polynomial or other basis functions, etc. We want to solve the following problem:
\begin{equation}
\theta^\star\in \arg\min_\theta \left(r(\theta) + \max_{x\in\XX} |f(x)-\hat f(x;\theta)|\right)
\label{eq:min-max-XX}
\end{equation}
where $r(\theta)$ is a regularization term, such as an $\ell_2$ or $\ell_1$ regularizer, or a penalty introduced 
to promote certain properties that $\hat f$ should have.

Solving Problem~\eqref{eq:min-max-XX} poses two main difficulties: ($i$) it is a bilevel optimization problem, due to the inner maximization problem over $\XX$; and ($ii$) the inner objective function is non-smooth due to the maximum operator,
which makes applying very efficient quasi-Newton methods, such as L-BFGS~\cite{LN89} 
more difficult.

To address the first issue, we consider an initial training dataset $(x_1,y_1)$, $\ldots$, $(x_{N_0},y_{N_0})$ of 
samples, where $x_k\in\XX$ and $y_k=f(x_k)$, and replace $\XX$ with $\{x_k\}_{k=1}^{N_0}$, so that 
solving the inner maximization problem simply amounts to finding the maximum of a finite number of terms. Accordingly, Problem~\eqref{eq:min-max-XX} is replaced by
\begin{equation}
    \theta^\star\in\arg\min_\theta \left(r(\theta) + \max_{k=1,\ldots,N_0} |f(x_k)-\hat f(x_k;\theta)|\right).
    \label{eq:min-max-k}
\end{equation}
A relevant problem addressed in this paper is how to actively learn a minimal number of samples $x_k$ so that the solution of~\eqref{eq:min-max-k} provides a solution to~\eqref{eq:min-max-XX}. 

We address the second issue by using the following smooth approximation of the max operator (cf.~\cite{Nes05}):

\begin{equation}
    \theta^\star\!\in\!\arg\min_\theta\! \big(r(\theta) + \frac{1}{\gamma}\!\log\!\big(\!\sum_{k=1}^{N_0}\big( 
    e^{\gamma(y_k-\hat f(x_k;\theta))}+e^{\gamma(\hat f(x_k;\theta)-y_k)}\big)\big)\big).  
\label{eq:softmax}%
\end{equation}

\begin{proposition}
\label{prop:softmax}
The minimization problem~\eqref{eq:softmax} has the same set of minimizers as~\eqref{eq:min-max-k}
in the limit $\gamma\to +\infty$.
\end{proposition}
\proof See Appendix~\ref{app:softmax}.\cvd

After obtaining the model $\hat f(\cdot;\theta^\star)$ by solving~\eqref{eq:softmax}, we can compute the WCE $\bar e^\star$ 
over the entire set $\XX$ by solving the following problem
\begin{equation}
    \bar e^\star = \max_{x\in\XX} |f(x)-\hat f(x;\theta^\star)|
    \label{eq:glob-opt}
\end{equation}
by using derivative-free global optimization methods~\cite{RS13,JSW98}. 

The main idea for enriching the dataset used in problem~\eqref{eq:min-max-k} is to add the obtained maximizer $x^\star$ and the corresponding value $y^\star=f(x^\star)$ to the training dataset and repeat the procedure until the maximum error $\bar e^\star$ is below a given threshold $\epsilon_{\rm err}$, or a preallocated computational budget is exhausted. In the latter case, as the WCE may not decrease at each iteration,
the model $\hat f(x;\theta)$ that has provided the smallest WCE during the iterations is selected as the final model.
The overall procedure is summarized in Algorithm~\ref{algo:max_error_fit}.

As in most active-learning schemes, we start from an initial number $N_0$ of samples $x_k$ purely based on their location in the input space, such as randomly from the uniform distribution over $\XX$, or by using Latin hypercube sampling (LHS)~\cite{MBC79}, or by gridding $\XX$ uniformly. 
Note that as $N_0$ increases, fewer active-learning steps are expected to be required to achieve a given WCE $\bar e^\star \le \epsilon_{\rm err}$, although the training problem \eqref{eq:softmax} becomes more complex at each step. 

\begin{algorithm}[h!]
    \caption{Worst-case nonlinear regression with active learning}
    \label{algo:max_error_fit}
    ~~\textbf{Input}: Function $f$ to approximate, model class $\FF$,
    number $N_0$ of initial samples to collect, maximum number $M$ of active-learning steps, 
    desired error threshold $\epsilon_{\rm err}$,
    smoothness parameter $\gamma$, regularization function $r$, input set $\XX$.
    \vspace*{.1cm}\hrule\vspace*{.1cm}
    \begin{enumerate}[label*=\arabic*., ref=\theenumi{}]
        \item Generate initial dataset $(x_1,y_1),\ldots,(x_{N_0},y_{N_0})$, $y_k=f(x_k)$;
        \item \textbf{Set} $N\leftarrow N_0$;
        \item \textbf{Repeat}:\label{line:repeat}
        \begin{enumerate}[label=\theenumi{}.\arabic*., ref=\theenumi{}.\arabic*]
            \item Solve problem~\eqref{eq:softmax} and get optimal model parameters $\theta_{N}$;
            \label{line:train-model}
            \item Set $x_{N+1}$ $\leftarrow$ global optimizer of problem~\eqref{eq:glob-opt};
            \label{line:glob-opt}
            \item Set $y_{N+1}\leftarrow f(x_{N+1})$, $e_{N}\leftarrow |y_{N+1}-\hat f(x_{N+1};\theta_N)|$;\label{line:add-sample}
            \item $N\leftarrow N+1$;\label{line:inc-N}
        \end{enumerate}
        \item \textbf{Until} $e_{N-1}\leq \epsilon_{\rm err}$ or $N=N_0+M$; 
        \label{line:stop-criteria}
        \item \textbf{Set} $i^\star\leftarrow \arg\min_{k=N_0,\ldots,N-1} e_k$,
        $\bar e^\star\leftarrow e_{i^\star}$, and $\theta^\star\leftarrow\theta_{i^\star}$;
        \label{line:best-error}
        \item \textbf{End}.
    \end{enumerate}
    \vspace*{.1cm}\hrule\vspace*{.1cm}
    ~~\textbf{Output}: Predictor $\hat f(x;\theta^\star)$ and error bound $\bar e^\star=\max_{x\in\XX}|f(x)-\hat f(x;\theta^\star)|$.
\end{algorithm}

Note that the objective function in~\eqref{eq:min-max-k} can be generalized to
the following MSE-regularized objective:
\begin{equation}
r(\theta) + \max_{k} \lvert f(x_k)-\hat f(x_k;\theta)\rvert + \frac{\nu}{N}\sum_{k=1}^N \bigl(f(x_k)-\hat f(x_k;\theta)\bigr)^2
\label{eq:nu-weight}
\end{equation}
where $\nu$ is a small positive scalar weight that discourages overfitting the model to the actively-learned corner-case samples. 

\subsection{Imposing a functional form on a set}
\label{sec:imposing}
Assume that we want to impose that $\hat f(x;\theta)\equiv w(x)$ for all $x$ in a given set $\GG=\{x:\ g(x)\leq 0,\ h(x)=0\}$, where $w:\XX\to\rr$, $g:\rr^n\to\rr^{n_g}$, and $h:\rr^n\to\rr^{n_h}$ are given functions. In Section~\ref{sec:mpqp} we will see an example with $w$ affine and $\GG$ polyhedral. 
Consider the indicator function $\delta:\rr^n\to\{0,1\}$ of $\GG$, i.e., $\delta(x)=1$ if $x\in\GG$ and $\delta(x)=0$ otherwise,
and either its piecewise affine (PWA) approximation
\begin{subequations}
\begin{equation}
    \hat\delta(x;\beta) = \max\left(1-\beta \max(\col(g(x),\pm h(x),0)),0\right)
    \label{eq:delta-pwa}
\end{equation}
or its approximation 
\begin{equation}
    \hat\delta(x;\beta) = e^{{-\beta\Big(\sum_{i=1}^{n_g}\maxzero{g_i(x)} + \sum_{j=1}^{n_h}|h_j(x)|\Big)}}
    \label{eq:delta-smooth}
\end{equation}
    \label{eq:delta}%
\end{subequations}
where $\beta>0$. Clearly, $\hat\delta(x;\beta)=1$, $\forall x\in\GG$.
Moreover, for $x\not\in \GG$, $\lim_{\beta \rightarrow+\infty}\hat\delta(x;\beta)$ $=0$.
Then, we define the model
\begin{equation}
    \hat f(x;\theta)= \hat\delta(x;\beta) w(x) + (1-\hat\delta(x;\beta))\hat v(x;\theta)
\label{eq:forced_gain}
\end{equation}
where $\hat v\in\FF$. Note that this changes the model class to which $\hat f$ belongs. The parameter $\beta$ can be either fixed or trainable; in the latter case, we assume $\beta$ is a component of $\theta$.

\subsection{Saturation}
\label{sec:saturation}
Assume that we want to bound the output of the model in the hyper-box
$\ymin\leq y\leq \ymax$. For example, we may know that the codomain of $f$ is contained in such a hyper-box. To this end, we consider the saturation function
\begin{subequations}
\begin{equation}
    \sat(y;\ymin,\ymax) = \min(\max(y,\ymin),\ymax)
\label{eq:sat-hard}
\end{equation}
and its smooth approximation (cf.~\cite[Section 3.B]{Bem25}):
\begin{equation}
    \softsat_\eta(y;\ymin,\ymax) = \ymax+\frac{1}{\eta}\log\left(\frac{1+e^{-\eta(y-\ymin)}}{1+e^{-\eta(y-\ymax)}}\right)
\label{eq:smooth_sat}
\end{equation}
\label{eq:sat}%
\end{subequations}
(all arithmetic operations are component-wise) where $\eta$ is either a fixed or a further trainable parameter, $\eta>0$.

\begin{proposition}
\label{prop:softsat}
Let $\ymin<\ymax$ component-wise and $\eta>0$. Then, the function $\softsat_\eta$ in~\eqref{eq:smooth_sat} is (component-wise) strictly monotonically increasing with respect to $y$, continuous, satisfies $\ymin\leq \softsat_\eta(y;\ymin,\ymax)\leq \ymax$, tends to $\ymin$ for $y\to -\infty$ and to $\ymax$ for $y\to +\infty$, and is such that
$\softsat_\eta\left(\frac{\ymin+\ymax}{2};\ymin,\ymax\right)=\frac{\ymin+\ymax}{2}$, $\forall \eta>0$. Moreover, it approaches the saturation function $\sat(y;\ymin,\ymax)$ asymptotically as $\eta\to +\infty$.
\end{proposition}
\proof See Appendix~\ref{app:softsat}.\cvd

Finally, we define the model
\begin{equation}
    \hat f(x;\theta)= \hat s(\hat v(x;\theta);\ymin,\ymax)
\label{eq:model-sat}
\end{equation}
where $\hat v\in\FF$ and $\hat s$ is either $\sat$ in~\eqref{eq:sat-hard} or $\softsat_\eta$ in~\eqref{eq:smooth_sat}
(we assume that $\eta$ is a component of $\theta$ in case it is trainable).
Note that this also changes the model class $\hat f$ belongs to.
Saturation can also be combined with the approach detailed in 
Section~\ref{sec:imposing} by setting
$\hat f(x;\theta)= \hat s((1-\hat\delta(x;\beta))\hat v(x;\theta) +\hat\delta(x;\beta)w(x);\ymin,\ymax)$.

\subsection{Inexact global optimization}
The error bound $\bar e^\star$ returned by Algorithm~\ref{algo:max_error_fit} is guaranteed to be an upper bound under the assumption that the global optimization problem~\eqref{eq:glob-opt} is solved exactly. 
Under mild assumptions on $f$ and $\hat f$, the following proposition shows that we can still guarantee an error bound when running the \texttt{direct} global optimization algorithm~\cite{JSW98,JM21} for a sufficiently large but finite number of iterations $T$.

\begin{proposition}
\label{prop:direct}
Assume $f:\XX\to\rr$ and $\hat f(\cdot;\theta^\star):\XX\to\rr$ have Lipschitz constants $K_f>0$ and $K_{\hat f}>0$, respectively. For any given tolerance $\delta>0$, there exists $T$ large enough such that, after running the global optimizer algorithm~\cite{JSW98} on problem~\eqref{eq:glob-opt} for $T$ iterations, we get
\begin{equation}
    \max_{x\in\XX}|f(x)-\hat f(x;\theta^\star)|\leq \bar e_T + \delta
    \label{eq:direct-bound}
\end{equation}
where $\bar e_T$ is the best value returned after $T$ iterations.
\end{proposition}

\proof
Let $g:\XX\to\rr$, with $g(x)\eqdef|f(x)-\hat f(x;\theta^\star)|$. For any $x,z\in\XX$, we get
$|g(x)-g(z)| = \bigl||f(x)-\hat f(x;\theta^\star)|-|f(z)-\hat f(z;\theta^\star)|\bigr|$
$\leq |(f(x)-\hat f(x;\theta^\star))-(f(z)-\hat f(z;\theta^\star))|$
$\leq |f(x)-f(z)|+|\hat f(x;\theta^\star)-\hat f(z;\theta^\star)|$
$\leq (K_f+K_{\hat f})\|x-z\|_2$,
where the first inequality follows from the reverse triangle inequality, and so $g$ is Lipschitz continuous with constant $K_g\leq K_f+K_{\hat f}$.

Let $x^\star\in\arg\max_{x\in\XX}g(x)$ and set $\epsilon=\delta/K_g$ (if $K_g=0$ then $g$ is constant and the result is trivial, since $\bar e_T=\bar e^\star$). By the everywhere-dense sampling property of DIRECT~\cite{JM21}, the algorithm eventually samples a point $x_\tau$, $\tau\leq T$, with $\|x^\star-x_\tau\|_2\leq\epsilon$. Hence, $\bar e_T \geq g(x_\tau)$, and by Lipschitz continuity of $g$, $g(x_\tau) \geq g(x^\star) - K_g\|x^\star-x_\tau\|_2 \geq \bar e^\star - K_g\epsilon = \bar e^\star-\delta$. \cvd

Note that solving problem~\eqref{eq:softmax} (in general nonconvex) suboptimally only affects the quality of the model $\theta^\star$, not the validity of the error bound $\bar e^\star$.

\section{Input-dependent error bounds}
\label{sec:input-dependent-bounds}
Algorithm~\ref{algo:max_error_fit} provides the global, symmetric, and uniform bound $\bar e^\star$ for $f(x)-\hat f(x;\theta^\star)$ over $\XX$, where $\theta^\star$ collects the optimal model coefficients determined by Algorithm~\ref{algo:max_error_fit} (possibly including $\beta^\star$ from~\eqref{eq:forced_gain} and $\eta^\star$ from~\eqref{eq:model-sat}). Here, we want to obtain bounds in the more general and therefore less conservative form
\[
    -e^\star_{\rm min}(x)\leq  f(x)-\hat f(x;\theta^\star) \leq e^\star_{\rm max}(x),\ \forall x\in\XX.
\]

\subsection{Constant asymmetric error bounds}
\label{sec:asymm-const}
Asymmetric uniform bounds, i.e., lower and upper bounds that are constant over $\XX$, can be computed as
\begin{equation}
    \begin{aligned}
    e^\star_{\rm min}(x)\equiv\bar e^\star_{\rm min} &\eqdef \max_{x\in\XX} \left(\maxzero{\hat f(x;\theta^\star)-f(x)}\right)\\
    e^\star_{\rm max}(x)\equiv \bar e^\star_{\rm max} &\eqdef \max_{x\in\XX} \left(\maxzero{f(x)-\hat f(x;\theta^\star)}\right)
    \end{aligned}
    \label{eq:envelope-asymm-uniform}
\end{equation}
by global optimization similarly to~\eqref{eq:glob-opt}.
Clearly, these are tighter than the symmetric bound $\bar e^\star$ returned by
Algorithm~\ref{algo:max_error_fit}, since $-\bar e^\star\leq -\bar e^\star_{\rm min}\leq f(x)-\hat f(x;\theta^\star) \leq \bar e^\star_{\rm max}\leq \bar e^\star$. The result of Proposition~\ref{prop:direct} can be immediately extended to the bounds in~\eqref{eq:envelope-asymm-uniform}. 

\subsection{Symmetric input-dependent error bounds}
Consider a set $\EE$ of positive functions $\varepsilon:\rr^n\times\rr^{n_\psi}\to\rr$  parameterized
by a vector $\psi\in\rr^{n_\psi}$, $\varepsilon(x;\psi)>0$, $\forall x\in\XX$, $\forall \psi\in\rr^{n_\psi}$. 
We consider the problem of learning a function $\varepsilon$ that upper bounds $|f(x)-\hat f(x;\theta^\star)|$ as tightly as possible. 
To this end, we want to solve the following training problem
\begin{equation}
    \begin{aligned}
    \psi^\star = \arg\min_\psi & \rho_\psi(\psi)+\frac{1}{N}\sum_{k=1}^{N} \mu(\varepsilon(x_k;\psi))\\
    \textrm{s.t.}\ &\varepsilon(x_k;\psi)\geq |e_k|,\ k=1,\ldots,N
    \end{aligned}
\label{eq:envelope}
\end{equation}
where $e_k=y_k-\hat f(x_k;\theta^\star)$ is the prediction error at sample $k$,
$\mu:\rr_+\to\rr_+$ is a monotonically increasing function (e.g., $\mu(\varepsilon)=\alpha\varepsilon$,
$\mu(\varepsilon)=\alpha\varepsilon^2$) that attempts squeezing the interval $[-\varepsilon(x_k;\psi),\varepsilon(x_k;\psi)]$ 
as much as possible, under the regularization $\rho_\psi$ on the parameters of function $\varepsilon$,
and $N\leq N_0+M$ is the number of available samples collected by Algorithm~\ref{algo:max_error_fit}.
We call $\varepsilon$ an {\it envelope} function, due to the constraint in~\eqref{eq:envelope}.

To relax the problem into an unconstrained nonlinear program, we rewrite the constraints in~\eqref{eq:envelope} into their equivalent form
\begin{equation}
    \max\left(0,\max_{k=1,\ldots,N}\left(|e_k|-\varepsilon(x_k;\psi)\right)\right) = 0
\label{eq:envelope2}
\end{equation}
and, by introducing an approximation of the maximum violation constraint in~\eqref{eq:envelope2}
similar to the one used in~\eqref{eq:softmax}, 
we relax~\eqref{eq:envelope2} into the following unconstrained nonlinear programming problem
\begin{equation}
    \begin{aligned}
    \min_\psi     
        \rho_\psi(\psi)\!+\!\frac{1}{N}\!\sum_{k=1}^{N} \mu(\varepsilon(x_k;\psi))
        \!+\!\frac{1}{\gamma_\varepsilon}\log\!\Big(1\!+\!\sum_{k=1}^{N} e^{\gamma_\varepsilon(|e_k|-\varepsilon(x_k;\psi))}\!\Big).
    \end{aligned}
\label{eq:loss-sigma}
\end{equation}
Note that, by Proposition~\ref{prop:softmax}, the last term in~\eqref{eq:loss-sigma} enforces constraint~\eqref{eq:envelope2} in the limit $\gamma_\varepsilon\to +\infty$, being a smooth version of $\max(0,\max_{k=1,\ldots,N}(|e_k|-\varepsilon(x_k;\psi)))$.

Any positive model $\varepsilon(x;\psi)$ can be used to learn the envelope functions. In this paper, we focus on NN models with two hidden layers  
\begin{equation}
    \begin{aligned}
    \varepsilon(x;\psi) = &W_3 a^+(W_2\cdot  a(W_1\cdot  x+b_1) + b_2)+b_3\\
    &W_3\geq 0,\ b_3>0
    \end{aligned}
\label{eq:varepsilon-NN}
\end{equation}
where $a^+$ is a nonnegative activation function, such as the ReLU or sigmoid function, 
$a:\rr^n\to\rr^m$ is any activation function, 
$W_1\in\rr^{n_1\times n}$, $W_2\in\rr^{n_2\times n_1}$, $W_3\in\rr^{1\times n_2}$ are weight matrices, and
$b_1\in\rr^{n_1}$, $b_2\in\rr^{n_2}$, $b_3\in\rr$ are bias terms. The transformation $a^+$ and the component-wise constraints on $W_3,b_3$ ensure that $\varepsilon(x;\psi)>0$ for all $x\in\rr^n$, with $\psi=\col(W_1,W_2,W_3,b_1,b_2,b_3)$. In particular, by setting $a^+$ and $a$ as ReLU functions we get PWA envelopes $\varepsilon$.

To get a global upper-bound function on $\XX$, we compute the following scaling factor $\kappa$ via global optimization:
\begin{equation}
    \kappa^\star = \max_{x\in\XX} \left(\frac{|f(x)-\hat f(x;\theta^\star)|}{\varepsilon(x;\psi^\star)}\right)
\label{eq:kappa-symm}
\end{equation}
and define the bounding functions
\begin{equation}
    \bar e^\star_{\rm min}(x) = \bar e^\star_{\rm max}(x) = \min(\bar e^\star,\kappa^\star \varepsilon(x;\psi^\star)).
    \label{eq:envelope-symm}
\end{equation}
Note that~\eqref{eq:kappa-symm} is closely related to split conformal prediction~\cite{AB23}, where the scaling factor $\kappa^\star$ is determined on a 
calibration dataset rather than optimizing over the entire set $\XX$, both for constant (homoscedastic) and input-dependent (heteroscedastic) bounds. In the latter case, our approach can also leverage additional calibration data when fitting $\varepsilon(x;\psi)$, so to limit conservativeness when making the envelope function valid over the entire $\XX$ in~\eqref{eq:envelope-symm}.

\subsection{Asymmetric input-dependent error bounds}
\label{sec:asymm-input-bounds}
To further reduce the conservativeness of the bounds, the approach described above can be extended to fit {\it asymmetric} 
input-dependent upper and lower error bounds. Let us consider two different envelope functions $\varepsilon(x;\psi_u)$ and $\varepsilon(x;\psi_\ell)$, with 
$\psi_u,\psi_\ell\in\rr^{n_\psi}$, to learn simultaneously by minimizing
\begin{equation}
    \begin{aligned}
    (\psi_u^\star,\psi_\ell^\star) = \arg\min & \rho_{\psi}(\psi_u)+\rho_{\psi}(\psi_\ell)\\
    &+\frac{1}{N}\sum_{k=1}^{N} \mu(\varepsilon(x_k;\psi_u)+\varepsilon(x_k;\psi_\ell))\\
    \textrm{s.t.}\ &\max\Big(0,\max_{k=1,\ldots,N}(e_k-\varepsilon(x_k;\psi_u)),\\
&\max_{k=1,\ldots,N}(-e_k-\varepsilon(x_k;\psi_\ell))\Big)=0.
\label{eq:envelope3}
    \end{aligned}
\end{equation}
Similarly to the envelope learning problem in~\eqref{eq:envelope}, we can relax the problem into the unconstrained nonlinear program
\begin{equation}
    \begin{aligned}
    \min_{\psi_u,\psi_\ell} &     
        \frac{1}{N}\sum_{k=1}^{N} \mu(\varepsilon(x_k;\psi_u)+\varepsilon(x_k;\psi_\ell))
        +\rho_\psi(\psi_u)+\rho_\psi(\psi_\ell)\\
        &\hspace*{-.8cm}+\frac{1}{\gamma_\varepsilon}\log\!\left(\!1+\sum_{k=1}^{N}\big( e^{\gamma_\varepsilon(e_k-\varepsilon^u(x_k;\psi_u))}+
        e^{\gamma_\varepsilon(-e_k-\varepsilon^\ell(x_k;\psi_\ell))}\big)\!\right)
    \label{eq:combined_upper_lower_training}
    \end{aligned}
\end{equation}
where the last term penalizes a smooth approximation of the maximum violation of the constraint in~\eqref{eq:envelope3}. Analogously to~\eqref{eq:glob-opt}, after finding the optimal parameters $\psi_u^\star$ and $\psi_\ell^\star$, we determine two scaling factors $\kappa_u^\star $ and $\kappa_\ell^\star $, respectively, by solving the global optimization problems
\begin{equation}
\kappa_u^\star = \max_{x\in\XX} \frac{\maxzero{f(x)-\hat f(x;\theta^\star)}}{\varepsilon(x;\psi_u^\star)},\
\kappa_\ell^\star = \max_{x\in\XX} \frac{\maxzero{\hat f(x;\theta^\star)-f(x)}}{\varepsilon(x;\psi_\ell^\star)}
\label{eq:kappa-asymm}
\end{equation}
to get the upper and lower bound functions on $\XX$
\begin{equation}
    \begin{aligned}
    \bar e^\star_{\rm min}(x) &=& \min(\bar e^\star\!\!,\kappa_\ell^\star\varepsilon^\ell(x;\psi^\star_\ell))\\
    \bar e^\star_{\rm max}(x) &=& \min(\bar e^\star\!\!,\kappa_u^\star\varepsilon^u(x;\psi^\star_u))
    \end{aligned}
    \label{eq:envelope-asymm}
\end{equation}
Note that, by replacing $\mu(\varepsilon^u(x;\psi_u)+\varepsilon^\ell(x;\psi_\ell))$ with $\mu(\varepsilon^u(x;\psi_u))+\mu(\varepsilon^\ell(x;\psi_\ell))$, the envelope functions $\varepsilon^u$ and $\varepsilon^\ell$ can be learned independently. Note also that any function $\hat f_s$ of $x$ whose values lie in 
$[\hat f(x;\theta^\star)-\bar e^\star_{\rm min}(x),\hat f(x;\theta^\star)+\bar e^\star_{\rm max}(x)]$ for all $x\in\XX$ is a valid predictor of $f(x)$; in particular, the uncertainty interval can be symmetrized by simply redefining the centered predictor
\begin{subequations}
\begin{equation}
    \hat f^s(x)
     = 
    \hat f(x;\theta^\star)+\frac{\bar e^\star_{\rm max}(x)-\bar e^\star_{\rm min}(x)}{2}
\label{eq:symmetric-f}
\end{equation}
along with the symmetric error bounding function
\begin{equation}
    \bar e^{\star s}_{\rm min}(x)=\bar e^{\star s}_{\rm max}(x) = \frac{\bar e^\star_{\rm max}(x)+\bar e^\star_{\rm min}(x)}{2}
    \label{eq:symmetric-e}
\end{equation}
\label{eq:symmetric-bounds}%
\end{subequations}
although this changes the model class to which $\hat f$ belongs.

\begin{example}
\label{ex:example-scalar}
Consider the scalar function $f(x)=\big(\sin((x-\frac{x^2}{10}))+(\frac{x}{10})^3-\frac{4x}{10}\big)\frac{e^{-x}}{1+e^{-x}}$ with $x\in[-10,10]$. We want to find an approximation of $f$ within the very simple class $\FF$ of NNs with two layers of 2 and 1 neurons, respectively, and $\tanh$ activation function. We run Algorithm~\ref{algo:max_error_fit} with 
$N_0=20$, $M=30$, $\epsilon_{\rm err}=0.001$, $\gamma=10$, $\nu=0$, and the regularization function $r(\theta)=10^{-8}\|\theta\|^2$. Figure~\ref{fig:example-scalar-error} shows how the maximum prediction error $e_N=|f(x)-\hat f(x;\theta)|$ on collected training data decreases during the execution of Algorithm~\ref{algo:max_error_fit}, reaching its 
minimum after around $N=30$ samples. The resulting model $\hat f(x;\theta^\star)$ is shown in Figure~\ref{fig:example-scalar-bounds} along with the uniform and input-dependent bounds $\bar e^\star_{\rm min}(x)$ and $\bar e^\star_{\rm max}(x)$. The input-dependent bounds are also NNs in $\FF$, also trained with $\ell_2$-regularization $\rho_\psi(\psi)=10^{-6}\|\psi\|_2^2$, $\mu(\varepsilon)=10^{-3}\varepsilon^2$, $\gamma_\varepsilon=100$. 

\begin{figure}[t]
\begin{center}
\includegraphics[width=\hsize]{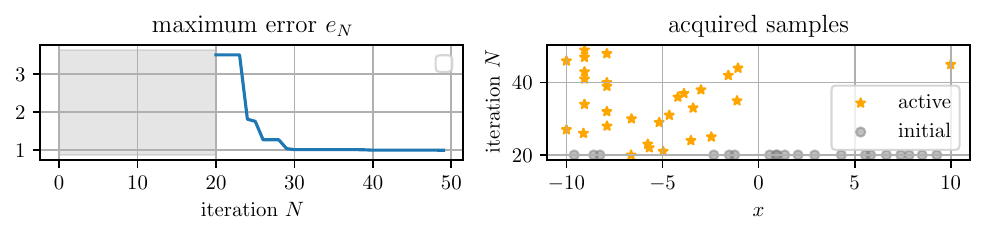}
\end{center}
\caption{Execution of Algorithm~\ref{algo:max_error_fit} on Example~\ref{ex:example-scalar}: maximum error $e_N$ 
(left plot), initialization (gray area), samples acquired (right plot). }
\label{fig:example-scalar-error}
\end{figure}

\begin{figure}[t]
\begin{center}
\includegraphics[width=\hsize]{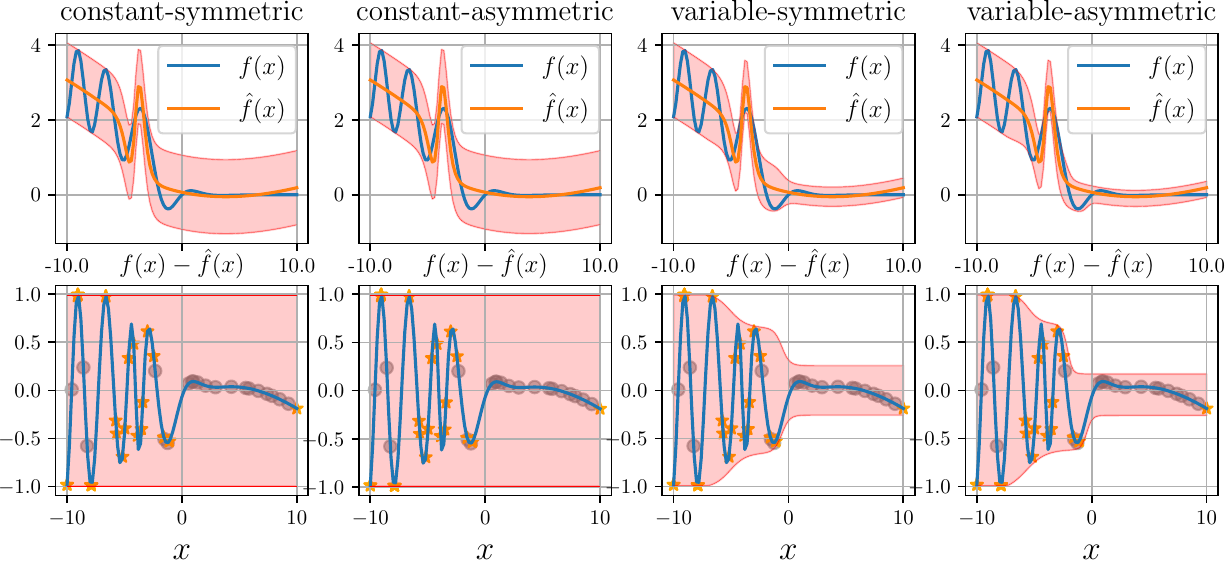}
\end{center}
\caption{Uncertainty bounds obtained on Example~\ref{ex:example-scalar}.}
\label{fig:example-scalar-bounds}
\end{figure}

\end{example}

\section{Constraint satisfaction}
\label{sec:constraint-satisfaction}
Assume that we are interested in replacing a given constraint $f(x)\leq 0$ with a sufficiently simpler constraint $\bar f(x)\leq 0$, such that 
\begin{equation}
    \bar f(x)\leq 0\ \implies\ f(x)\leq 0,\ \forall x\in\XX
\label{eq:bar-f}
\end{equation}
in the least conservative way, i.e., minimizing the size of the set $\{x\in\XX:\bar f(x)>0,\ f(x)\leq 0\}$ of feasible points for $f$ that are declared infeasible by $\bar f$. The actual values taken by $f(x)$ are not relevant
for~\eqref{eq:bar-f}, as long as the sign of $f(x)$ is correctly predicted by $\bar f(x)$, i.e.,
$\sign(\bar f(x)) =-1 \ \implies\ \sign(f(x)) = -1,\ \forall x\in\XX$.
Therefore, we apply the worst-case regression approach developed in Section~\ref{sec:worst-case-regression}
with $f(x)$ replaced by $s(f(x))$ and $\hat f(x;\theta)$ by $s(\hat f(x;\theta))$, where $s:\rr\to\rr$ is a smooth approximation of the sign function. A possible choice is $s(x)=\tanh(\eta x)$, where $\eta>0$ is a given parameter (e.g., $\eta=10$). Note that $s$ is only used during training and then removed after determining the optimal parameter vector $\theta^\star$.

After stopping the active-learning process and having fixed the optimal parameter vector $\theta^\star$, we 
need to make sure to construct $\bar f(x)$ such that it cannot be negative when $f(x)>0$, so as to guarantee that
condition~\eqref{eq:bar-f} is satisfied. To this end, we
compute the smallest value of $\hat f(x;\theta^\star)$ when the condition $f(x)>0$ (i.e., $\sign (f(x))=1$), by solving the following global optimization problem:
\begin{equation}
    \Delta f \eqdef  \min_{x\in\XX} \left(\frac{1}{2}(1+\sign(f(x)))\hat f(x;\theta^\star)\right)
    \label{eq:delta_f}
\end{equation}
and define $\bar f(x)$ as in the next proposition.

\begin{proposition}
\label{prop:constraint-satisfaction}
Let $\Delta f$ be the solution of~\eqref{eq:delta_f} and define
\begin{equation}
    \bar f(x) \eqdef \hat f(x;\theta^\star) - \Delta f+\epsilon_f
\label{eq:bar-f-e_star}
\end{equation}
for an arbitrarily small number $\epsilon_f>0$. Then, condition~\eqref{eq:bar-f} is satisfied.
\end{proposition}
\proof Assume, by contradiction, that there exists a vector $x$ such that $\bar f(x)\leq 0$ and $f(x)>0$. Then,
\[
\begin{aligned}
0&\geq \bar f(x) = \hat f(x;\theta^\star) - \Delta f+\epsilon_f\\
&\geq \hat f(x;\theta^\star) - \hat f(x;\theta^\star) + \epsilon_f = \epsilon_f
\end{aligned}
\]
which contradicts the assumption $\epsilon_f>0$. Hence,~\eqref{eq:bar-f} must hold.
\cvd

Note that, by selecting $\hat f$ as the composition of a strictly-increasing function $\hat f_m:\rr\to\rr$ 
with another function $\hat f_c:\rr^n\to\rr$, parameterized by $\theta_m$ and $\theta_c$, respectively, $\hat f(x;\theta)=\hat f_m(\hat f_c(x;\theta_c);\theta_m)$, the resulting constraint~\eqref{eq:bar-f-e_star} 
is equivalent to the constraint
\begin{equation}
    \hat f_c(x;\theta_c^\star)\leq \hat f_m^{-1}\left(\Delta f-\epsilon_f;\theta_m^\star\right).
\label{eq:f_c-constraint}
\end{equation}
In particular, if $\hat f_c$ is convex, such as an input-convex NN~\cite{SBB25}, 
the constraint in~\eqref{eq:f_c-constraint} is convex. In the special case $\hat f_c=\max_{i=1,\ldots,n_f}(A_ix-b_i)$, we get the convex polyhedral set $\{x:\ Ax\leq b-(\epsilon_f-\Delta f)\mathbf{1}\}$.

\section{Modeling and Controller Approximations}
\subsection{Discrete-time uncertain models}
\label{sec:dyn-sys}
We are given the continuous-time nonlinear model of a dynamical system 
\begin{equation}
    \begin{aligned}
    \dot \xi(t) &=& F(\xi(t),u(t))\\
    \tau(t) &=& G(\xi(t),u(t))
    \end{aligned}
\label{eq:ct-model}
\end{equation}
where $\xi(t)\in\rr^{n_\xi}$ is the state vector and $u(t)\in\rr^{n_u}$ the control input vector,
derived, for example, from first-principles. Our goal is to learn a discrete-time uncertain model of the form
\begin{equation}
    \begin{aligned}
    \xi(t+T_s) &=& \hat F(\xi(t),u(t);\theta) + w(t)\\
    \tau(t) &=& \hat G(\xi(t),u(t);\theta) + v(t)
    \end{aligned}
\label{eq:discrete-time-model}
\end{equation}
where $T_s>0$ is the sampling time, $\hat F:\rr^{n_\xi}\times\rr^{n_u}\to\rr^{n_\xi}$ and $\hat G:\rr^{n_\xi}\times\rr^{n_u}\to\rr^{n_\tau}$ are models parameterized by $\theta$, and we treat the modeling errors $w(t)\in\rr^{n_\xi}$ and $v(t)\in\rr^{n_\tau}$ as unknown disturbance vectors affecting the state update and output equations, respectively, $w(t)\in\WW$ and $v(t)\in\VV$, where $\WW\subset\rr^{n_\xi}$ and $\VV\subset\rr^{n_\tau}$ are unknown hyper-boxes.
Our goal is to learn $\hat F,\hat G$ along with the disturbance sets $\WW,\VV$. For example, if $\hat F, \hat G$ are linear models, we would get a conservative uncertain discrete-time linear state-space model of~\eqref{eq:ct-model} with unknown additive disturbances whose bounds are known. Similarly, if $\hat F, \hat G$ are NNs with ReLU activation functions, we would get an uncertain PWA discrete-time model of~\eqref{eq:ct-model}.

We can solve the above problem by applying the worst-case regression approach of Section~\ref{sec:worst-case-regression} component-wise to the functions $F$ and $G$, with $x=\col(\xi(t),$ $u(t))$ and
$y=f(x)=\xi_j(t+T_s)$ for each state $j=1,\ldots,n_\xi$, and $y=f(x)=\tau_j(t)$ for each output $j=1,\ldots,n_\tau$. Note that evaluating $\xi(t+T_s)$ requires integrating the continuous-time model~\eqref{eq:ct-model} from $t$ to $t+T_s$ with initial condition $\xi(t)$ and constant input $u(t)$. The uncertainty bounds for each component of $F$ and $G$ can then be obtained by solving~\eqref{eq:envelope-asymm-uniform} and combined to form the hyper-boxes $\WW$ and $\VV$.

\subsection{Approximate mpQP and explicit linear MPC}
\label{sec:mpqp}
Multiparametric quadratic programming (mpQP) problems~\cite{BMDP02a}
\begin{equation}
    \begin{aligned}
        z^\star(x)= &\arg\min_z \frac{1}{2}z'Q z + (Fx+f)'z \\
        & \text{s.t. } Az\leq Bx+b,\quad x\in\XX
    \end{aligned}
\label{eq:mpQP}
\end{equation}
arise in explicit model predictive control (MPC) of linear time-invariant systems~\cite{Bem21c}. 
The solution $z^\star(x)$ is a PWA function of the parameter vector $x\in\rr^{n_x}$ over a polyhedral partition of the parameter set $\XX\subseteq\rr^{n_x}$ in $N_r$ regions, and can be computed offline~\cite{BMDP02a}. As the number of regions $N_r$ grows considerably with the number of constraints, we want to learn a surrogate model $\hat f$ of 
selected components $f(x)\eqdef[I\ 0\ \ldots\ 0]z^\star(x)$ of the solution $z^\star$ over $\XX$.

A typical reference-tracking formulation used in applications 
is
\begin{equation}
\begin{aligned}
        \min_{z} \sum_{k=0}^{N-1} &(\tau_{k+1}-r)'Q_\tau(\tau_{k+1}-r) + \Delta u_k'Q_{\Delta u}\Delta u_k
        +\rho_2\eps^2+\rho_1\eps\\   
        \st & \xi_{k+1} = \bar A\xi_k + \bar B u_k,\ k=0,\ldots,N-1\\
        & \tau_k = \bar C\xi_k,\ k=1,\ldots,N\\
        & u_{k} = u_{k-1}+\Delta u_k,\ k=0,\ldots,N-1\\
        & \umin \leq u_k \leq \umax,\ k=0,\ldots,N_u-1\\
        & \dumin \leq \Delta u_k \leq \dumax,\ k=0,\ldots,N_u-1\\
        & \tau_{\rm min}-V_{\rm min}\eps \leq \tau_k \leq \tau_{\rm max}+V_{\rm max}\eps,\ k=1,\ldots,N_c\\
        & \Delta u_k = 0,\ k=N_u,\ldots,N-1,\ \eps \geq 0
\end{aligned}
\label{eq:linear-MPC}
\end{equation}
where $k$ is the prediction step, 
$\Delta u_k\in\rr^{n_u}$ are the input increments, $(\bar A, \bar B, \bar C)$ is a state-space realization of the model of the controlled process
$\xi(t+1) = \bar A\xi(t)+ \bar B u(t)$, $\tau(t) = \bar C\xi(t)$,
matrix $Q_\tau$ is the weight on tracking errors, 
$Q_\tau=Q_\tau'\succeq 0$, $Q_\tau\in\rr^{n_\tau\times n_\tau}$ and
$Q_{\Delta u}$ the weight of control increments, $Q_{\Delta u}=Q_{\Delta u}'\succ 0$, 
$Q_{\Delta u}\in\rr^{n_u\times n_u}$;
$N$ the prediction horizon, $N_u$ the number of free control moves, $N_c$ the number of output constraints in prediction; $\zeta\in\rr$ is a nonnegative slack variable used to soften output constraints via the given ECR (equal concern relaxation) nonnegative vectors $V_{\rm min},V_{\rm max}\in\rr^{n_\tau}$,
and is penalized with weights $\rho_2,\rho_1\geq 0$. 

Problem~\eqref{eq:linear-MPC} can be transformed into the mpQP~\eqref{eq:mpQP} by defining 
$x=\col(\xi_0,r$, $u_{-1})$, where $\xi_0=\xi(t)\in\rr^{n_\xi}$ is the current state vector, 
$r=r(t)\in\rr^{n_\tau}$ is the current output reference value, and $u_{-1}=u(t-1)\in\rr^{n_u}$ is the last control input applied to the process, $z=\col(u_0,\ldots,u_{N_u-1},\zeta)$ is the sequence of future control moves optimized over the prediction horizon and slack, and $y=f(x)$ is the optimal value of $u_0$, i.e., the control input $u(t)$ that gets applied to the process.

Let us assume $n_u=1$ for simplicity of notation; in the case $n_u>1$ we treat the approximation problem component-wise, i.e., we learn a surrogate model $\hat f_i$ for each component $i=1,\ldots,n_u$ of the optimal solution $f(x)$, using the approaches developed in Section~\ref{sec:worst-case-regression}.

Assuming $\umin\leq\umax$, $\dumin\leq\dumax$, Problem~\eqref{eq:mpQP} is feasible for all $x\in\XX$ and such that the matrix $Q$ is positive definite, so that the solution $z^\star(x)$ exists and is unique.
When no constraints are active at the optimal solution $z^\star(x)$, we have that $f(x)$ coincides
with the unconstrained solution
\begin{equation}
    w(x)=-[1\ 0\ \ldots\ 0]Q^{-1}(Fx+f)
    \label{eq:unconstrained-QP}
\end{equation}
for all and only the values of $x$ satisfying
\begin{equation}
    H_0x\leq K_0,\quad H_0\eqdef -AQ^{-1}F-B,\ K_0\eqdef b+AQ^{-1}f.
\label{eq:CR0}
\end{equation}
As we wish the surrogate model $\hat f$ to coincide with the unconstrained solution~\eqref{eq:unconstrained-QP} 
when the active set is empty, we consider the approximations of the indicator function defined in~\eqref{eq:delta}
with $g(x)=\bar H_0x-\bar K_0$ and $n_h=0$, where $\{x:\ \bar H_0x\leq \bar K_0\}$ is a minimal 
hyperplane representation of $\{x:\ H_0 x\leq K_0\}$. 
Moreover, to enforce the constraints on the first control input $u_0$,
we also consider the saturation functions~\eqref{eq:sat}
with $\ymin=\max(u_{{\rm min}}, u_{-1}+\dumin)$, $\ymax=\min(u_{{\rm max}}, u_{-1}+\dumax)$. 

The approximation error can be seen as an input disturbance $\delta_u\in\rr^{n_u}$, $|\delta_u|\leq \bar e^\star$, affecting the control input $u(t)$ applied to the process. Hence, a possible use of Algorithm~\ref{algo:max_error_fit} is to start from 
a robust MPC design with given uncertainty $|\delta_u|\leq \bar\delta_u$, evaluate its approximation,
and verify that $\bar e^\star\leq \bar\delta_u$.

\section{Numerical examples}
All numerical tests were run in Python 3.11 with the \texttt{maxfit} package on a MacBook Pro with Apple M4 Max (16 CPU cores), using \texttt{direct}~\cite{JSW98,JM21} for global optimization (absolute tolerance $10^{-8}$, relative tolerance $10^{-5}$, max 2000 evaluations). Unless otherwise specified, $\gamma=10$ and $r(\theta)=10^{-8}\|\theta\|^2$.

\subsection{Nonlinear function}
We want to approximate the Gaussian function $f(x)=e^{-30((x_1-0.5)^2+(x_2-0.5)^2)}$ on the box $\XX=[0,1]^2$
with a PWA one and quantify the WCE. We run Algorithm~\ref{algo:max_error_fit} with an initial dataset of $N_0=100$ training points, for $M=50$ active-learning steps. We consider the class $\FF$ of NN models $\NN(x;\theta)$ with two layers of 10 and 5 neurons, respectively, 
while for the input-dependent bound functions $\NN(x;\psi)$ we consider two layers of 20 and 10 neurons, respectively, with leaky-ReLU activation function $a(x)=\max(x,0.1x)$ in both. The CPU time is $114.3$\,s for running Algorithm~\ref{algo:max_error_fit} (with $106.0$\,s spent on training the surrogate and $7.8$\,s in global optimization) and $6.4$\,s to compute an asymmetric input-dependent confidence interval as described in Section~\ref{sec:asymm-input-bounds}, with $\rho_\psi(\psi)=10^{-8}\|\psi\|^2$. 

\begin{figure}[t]
\begin{center}  
\includegraphics[width=.48\hsize]{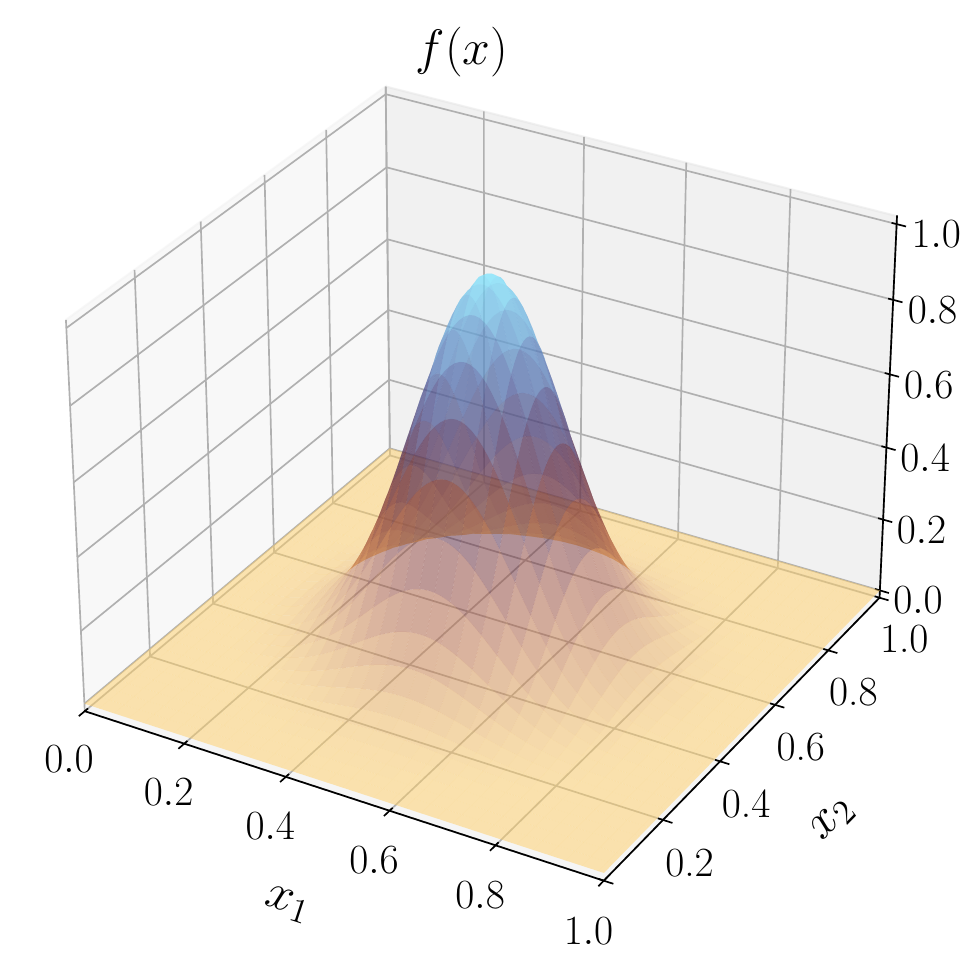}~\includegraphics[width=.48\hsize]{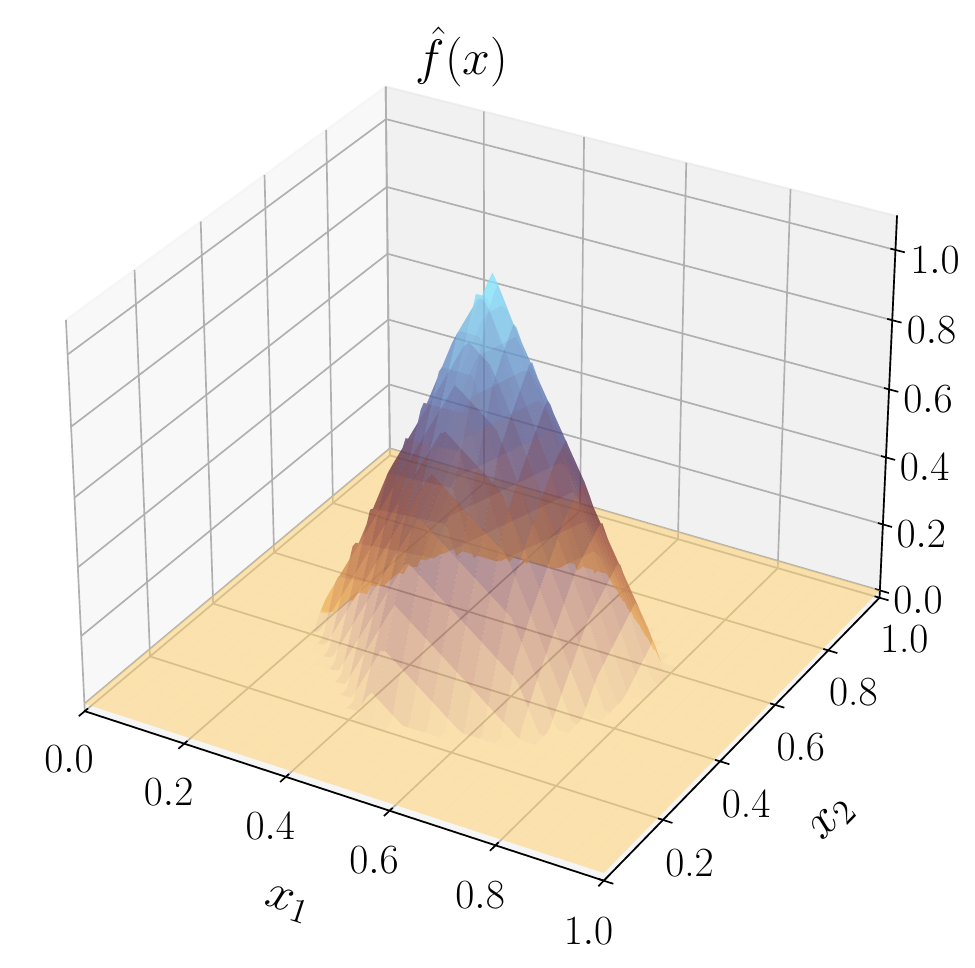}\\
\includegraphics[width=.48\hsize]{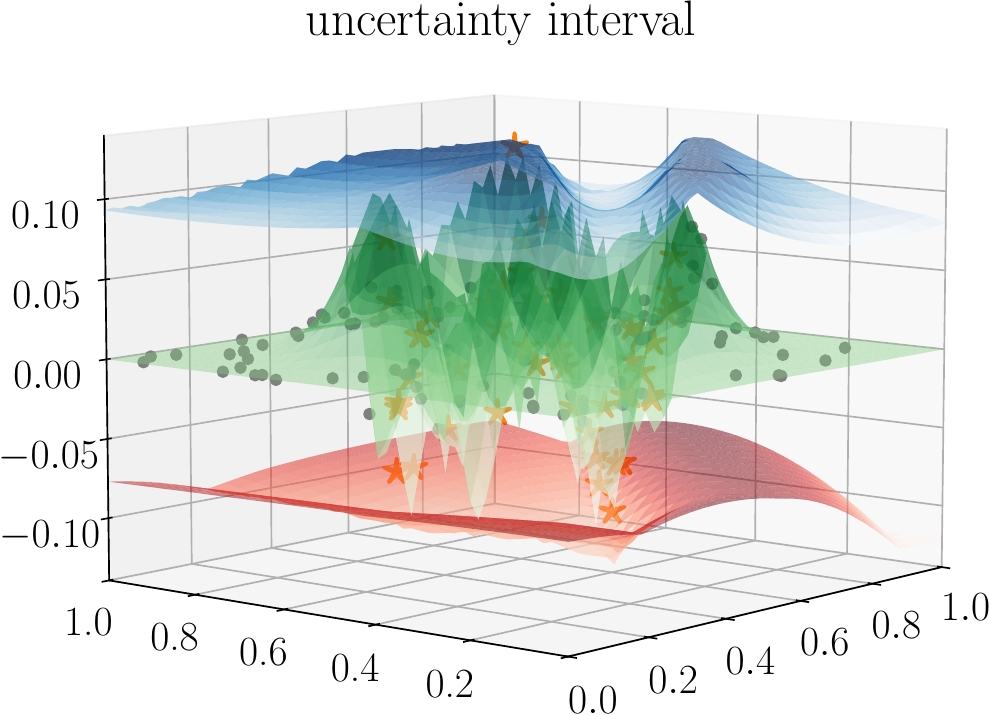}~
\includegraphics[width=.48\hsize]{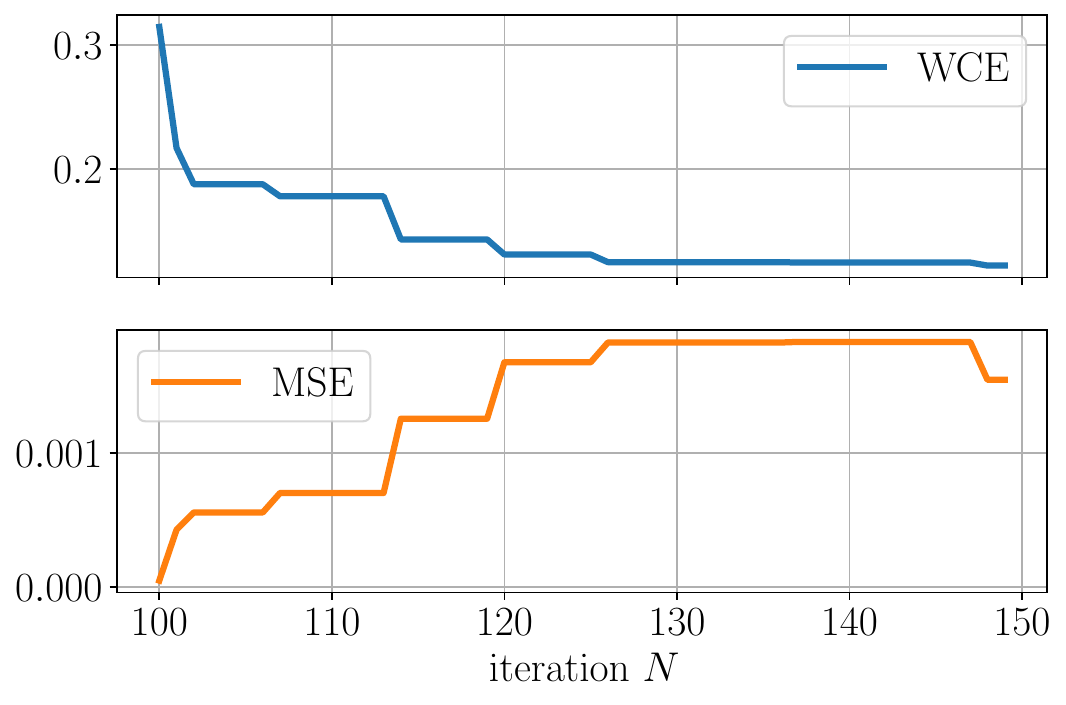}
\end{center}
\caption{Gaussian function. Top-left: function $f(x)$; Top-right: learned leaky-ReLU $\hat f(x;\theta)$; Bottom-left: pointwise error $e(x)=f(x)-\hat f(x;\theta)$, envelope $[\bar e^\star_{\min}(x),\bar e^\star_{\max}(x)]$, actively-learned
samples (orange stars) and initial samples (gray dots); Bottom-right: 
WCE and MSE over active-learning iterations of Algorithm~\ref{algo:max_error_fit}.}
\label{fig:Gaussian-1}  
\end{figure}

The results are shown in Figure~\ref{fig:Gaussian-1}. 
Note that, although the MSE is considerably smaller than the WCE, minimizing the WCE, even if related (cf.~\cite[Th. 1.3]{Pow81}), does not minimize the MSE, as expected.

\subsection{Convex inner approximation of a nonconvex set}
\begin{figure}[t]
\begin{center}
\includegraphics[width=.45\hsize]{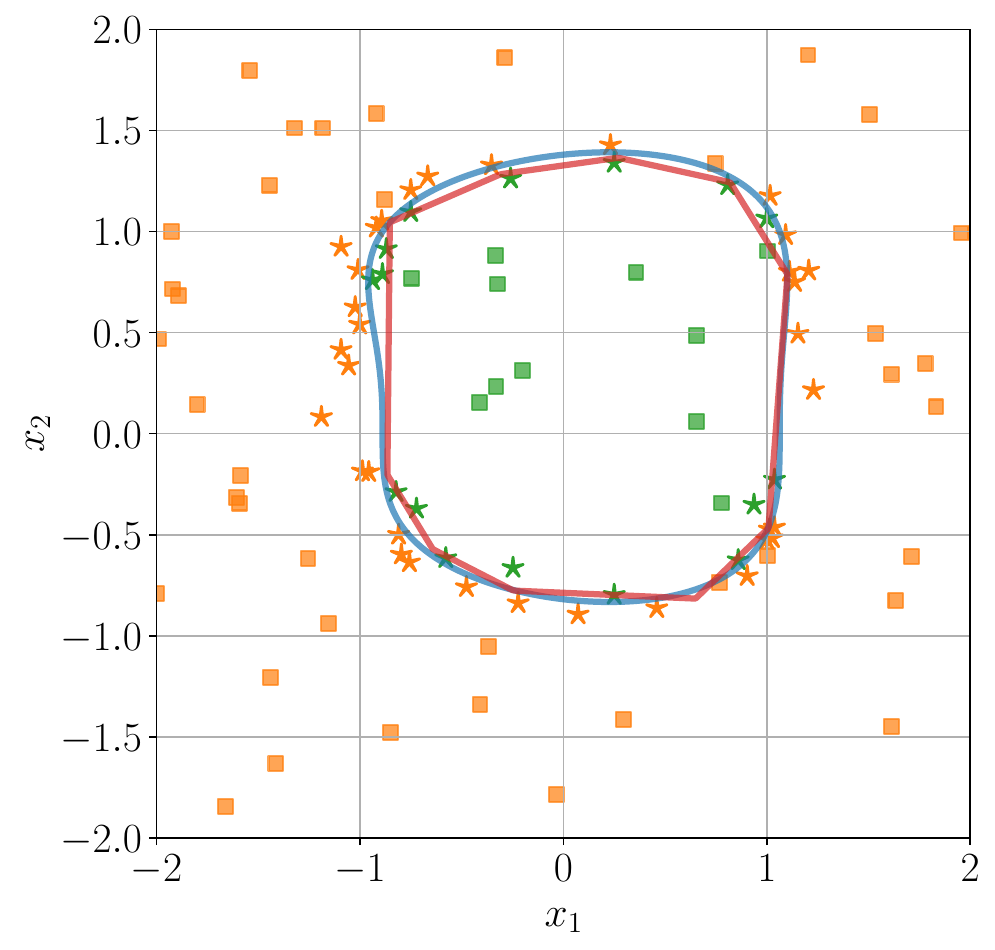}\hspace*{1cm}\includegraphics[width=.45\hsize]{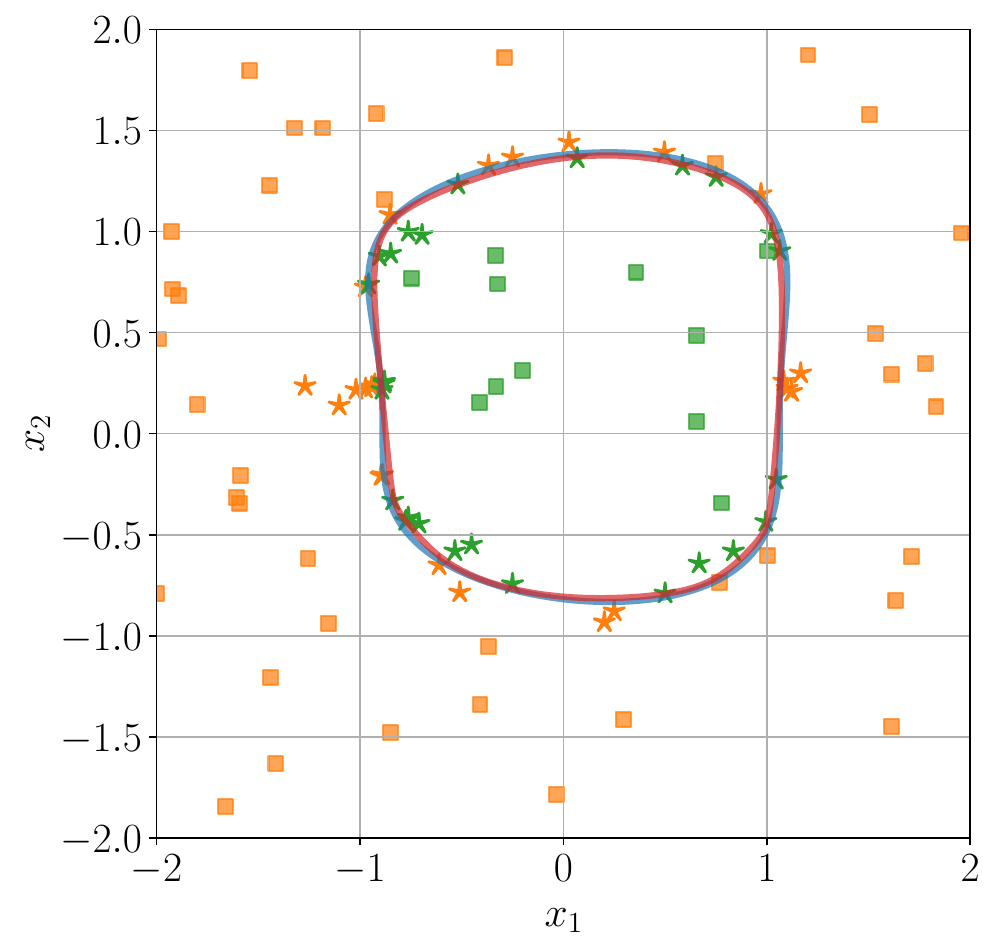}
\end{center}
\caption{Convex inner approximation of the nonconvex set $\Ss$ (blue): polyhedron (left) and input-convex NN (right). Feasible points (green), infeasible points (orange), initial training points (boxes), actively-learned points (stars).}
\label{fig:convex-inner-approx}
\end{figure}

We want to find a convex inner approximation of the following nonconvex set
\[
    \Ss = \{x\in\rr^2:\ x_1^2+x_2^4 + \frac{1}{3}x_1^3 - x_2^3-\frac{1}{2}x_2\leq 1\}
\]
with $x_1\in[-2,2]$, $x_2\in[-2,2]$. Following the approach described in Section~\ref{sec:constraint-satisfaction}, 
we run Algorithm~\ref{algo:max_error_fit} with  $N_0=50$ for $M=50$ active-learning steps, using 
$\eta=10$ and the regularization function $r(\theta)=10^{-4}\|\theta\|^2$. We consider two different model classes $\FF$ for $\hat f$: ($i$) a convex PWA function $\hat f(x;\theta)=\max_{i=1\ldots n_f}A_ix-b_i$, where $n_f=10$ and $\theta=\col(A_1^\top,\ldots,A_{n_f}^\top,b)$ collects the learnable parameters, which provides the convex polyhedral set $\{x\in\rr^2:\ Ax\leq b-\Delta f+\epsilon_f\}$; ($ii$) an input-convex NN with 2 hidden layers of 8 neurons each, linear bypass, and softplus activation function $a:\rr\to\rr$, $\hat f(x;\theta)=W_3^2a(W_2^2a(V_1x+b_1)+V_2x+b_2)+V_3x+b_3$, where $(\cdot)^2$ denotes component-wise square and ensures nonnegative weights and, therefore, convexity~\cite{SBB25}. 
The given set $\Ss$ and the convex inner approximations obtained with the two model classes are shown in Figure~\ref{fig:convex-inner-approx}. They are computed in $52.6$\,s
($45.7$\,s for training, $6.4$\,s for global optimization) 
and $156.8$~s ($146.3$\,s training / $9.6$\,s global optimization), respectively, with a
resulting $\Delta f = -0.07817$ (polyhedron) and $\Delta f = -0.03612$ (input-convex NN).
In both cases, by Proposition~\ref{prop:constraint-satisfaction}, any $x\in[-2,2]^2$ such that $\hat f(x;\theta)-\Delta f<0$ implies that $x\in\Ss$.

\subsection{Uncertain discrete-time model}
We consider the following continuous-time nonlinear model of a pendulum with friction and nonlinear stiffness:
\[
    \begin{aligned}
    \dot \xi_1(t) &= \xi_2(t)\\
    \dot \xi_2(t) &= \frac{1}{J}(u(t)-b\xi_2(t) - m g \ell_c \sin(\xi_1(t))- k_1 \xi_1(t) - k_3\xi_1(t)^3)
    \end{aligned}
\]
where $\xi_1(t)$ is the angular position (rad), $\xi_2(t)$ the angular velocity (rad/s), and $u(t)$ the applied torque (Nm), with parameters $J=0.05$~kgm$^2$, $b=0.08$~Nms/rad, $m=1$~kg, $g=9.81$~m/s$^2$, $\ell_c=0.15$~m, $k_1=2$~Nm/rad, $k_3=5.0$~Nm/rad$^3$. We want to learn a discrete-time uncertain model of the form~\eqref{eq:discrete-time-model} with sampling time $T_s=0.1$~s for the range $\xi_1(t)\in[-\pi,\pi]$~rad, $\xi_2(t)\in[-5,5]$~rad/s, $u(t)\in[-2,2]$~Nm. Algorithm~\ref{algo:max_error_fit} is run with $x=\col(\xi(t),u(t))$ starting from
$N_0=100$ samples, generated by LHS, for $M=50$ active-learning steps.

\begin{figure}
\begin{center}
\includegraphics[width=.45\hsize]{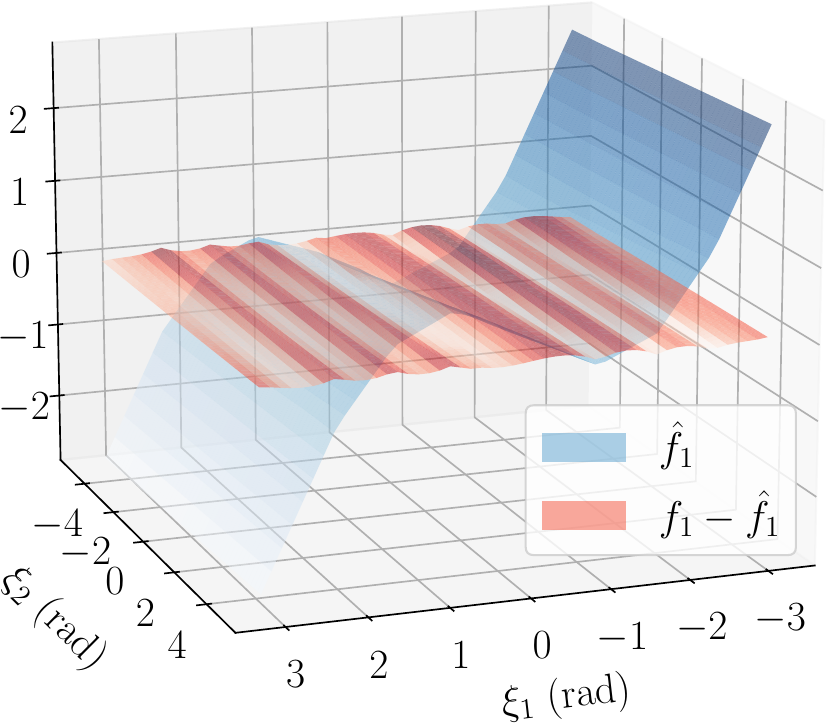}~\includegraphics[width=.45\hsize]{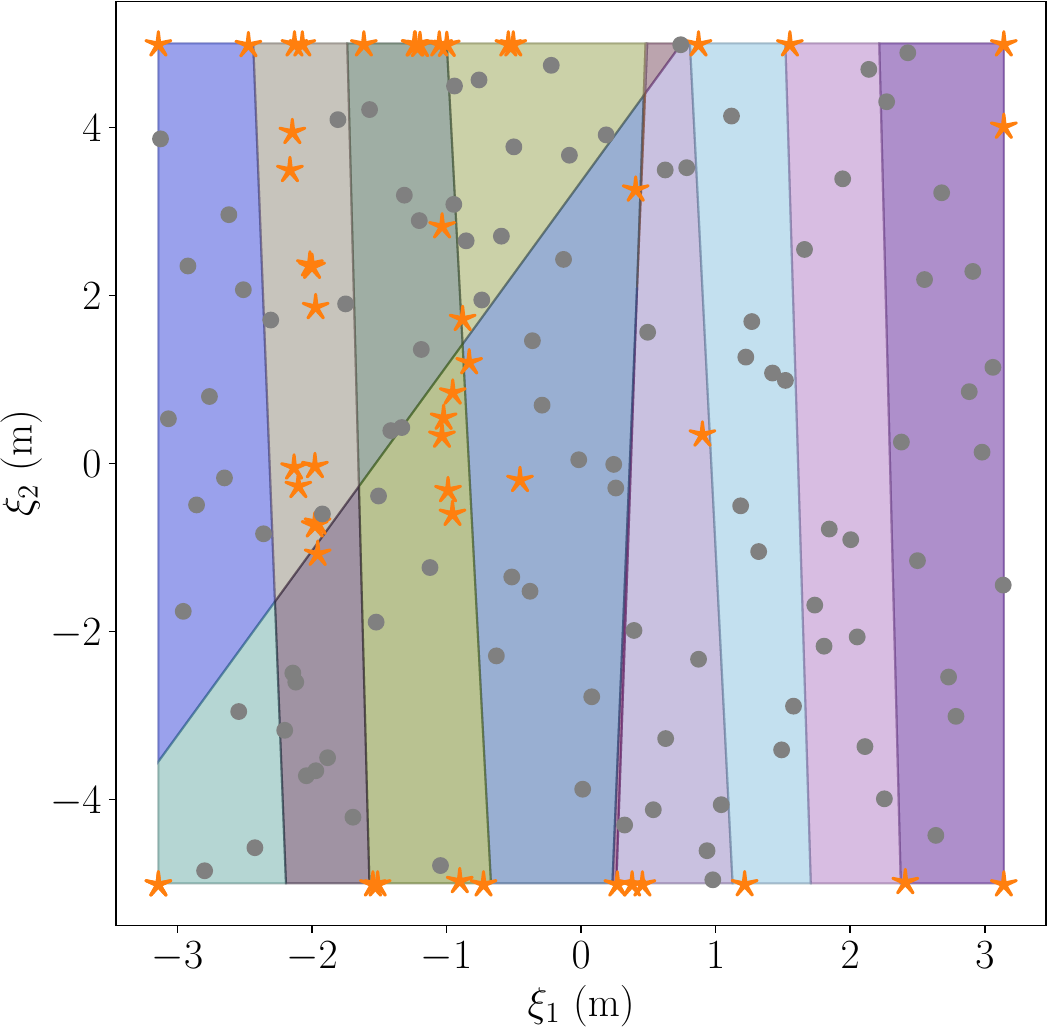}
\end{center}
\caption{Pendulum example: discrete-time ReLU mapping for $\xi_1(k+1)$ and approximation error (left), and regions of PWA equivalent of ReLU network (right) at $u=0$~Nm.}
\label{fig:pendulum-model}
\end{figure}

We consider the class $\FF$ of shallow NN models $\NN(x;\theta)$ with 10 neurons and ReLU activation function plus a linear function of $u$ to approximate the model given by integrating the continuous-time model from $t$ to $t+T_s$ with initial condition $\xi(t)$ and constant input $u(t)$ with Heun's method from the \texttt{diffrax} package~\cite{Kid21}. The results are shown in Figure~\ref{fig:pendulum-model} for the first state. The CPU time for running Algorithm~\ref{algo:max_error_fit} is $152.1$~s for the first state ($105.4\,s$ for training, $45.8$\,s for global optimization) and $176.3$~s for the second state ($118.4\,s$ for training, $72.2$\,s for global optimization). The resulting additive uncertainty bounds are $\WW=[-0.0810,0.0810]\times[-2.6368,2.6368]$,
corresponding to a maximum relative error of $2.82\%$ for the first state and $5.67\%$ for the second state.

\subsection{Multiparametric quadratic programming}
We consider a randomly-generated mpQP problem in the form~\eqref{eq:mpQP} with $x\in\rr^2$, $z\in\rr^{10}$, 30 linear inequality constraints, and bound constraints $-\zmin\leq z_i\leq \zmax$, $i=1,\ldots,10$, taking the first component $y=z_1$ of the optimal solution as the value to approximate. The exact mpQP solution $f(x)$, computed using the mpQP solver~\cite{pdaqp}, has 162 polyhedral regions. We consider the class $\FF$ of NN models $\NN(x;\theta)$ with $5$ neurons in each layer,  linear bypass in each layer, and ReLU activation function. In order to have that the unconstrained solution~\eqref{eq:unconstrained-QP} of the QP problem is exactly represented by $\hat f$ in the corresponding region~\eqref{eq:CR0}, we set
\begin{equation}
    \begin{aligned}
    \hat f(x;\theta) = &\min(\max(\hat\delta(x;\beta)\left(-[1\ 0\ \ldots\ 0]Q^{-1}(Fx+f)\right) \\
    &+ (1-\hat\delta(x;\beta))\NN(x;\theta),\zmin),\zmax)
    \end{aligned}
\label{eq:mpqp-network}
\end{equation}
where we used the saturation function $\sat$ as in~\eqref{eq:sat-hard} with $\zmin=-1$, $\zmax=1$,
and defined $\hat\delta(x;\beta)$ from~\eqref{eq:delta-pwa} as
$\hat\delta(x;\beta) = \max\left(1-\beta \max\left(H_0x-K_0,0\right),0\right)$, 
with $H_0,K_0$ as in~\eqref{eq:CR0} and $\beta$ is a trainable parameter.

\begin{figure}[t]
\begin{center}  
\includegraphics[width=.48\hsize]{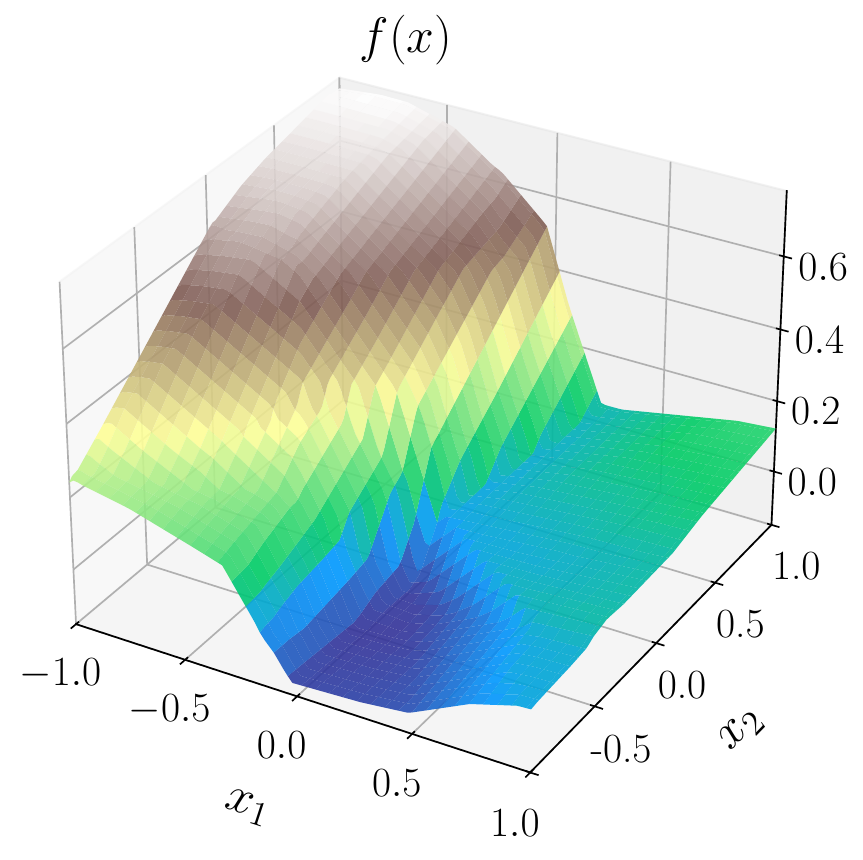}~\includegraphics[width=.48\hsize]{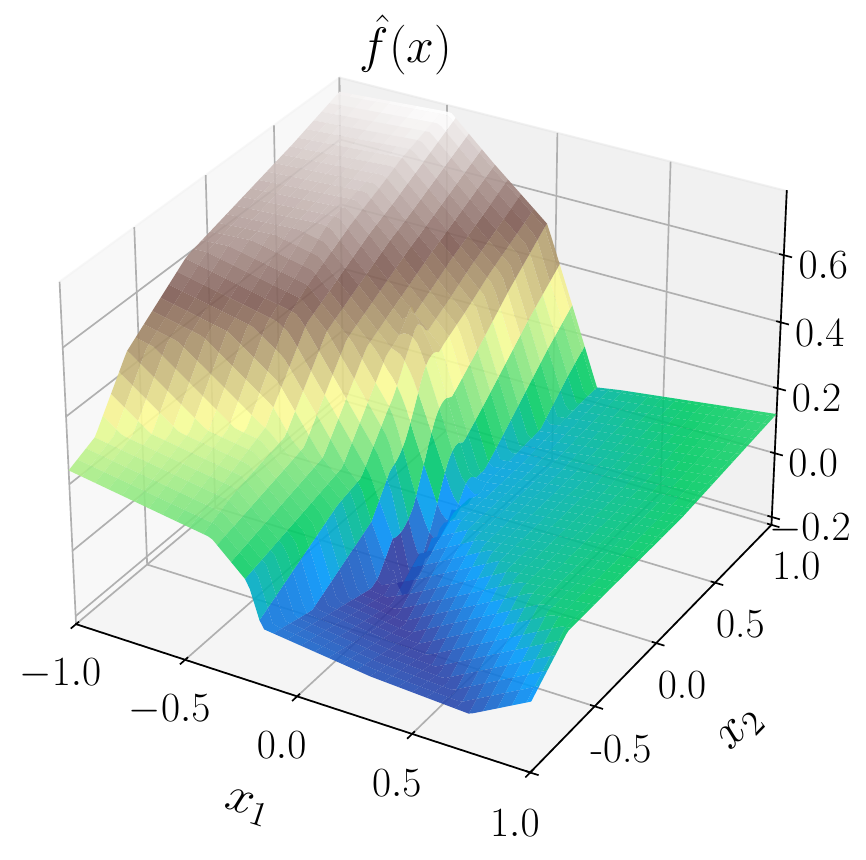}\\
\includegraphics[width=.48\hsize]{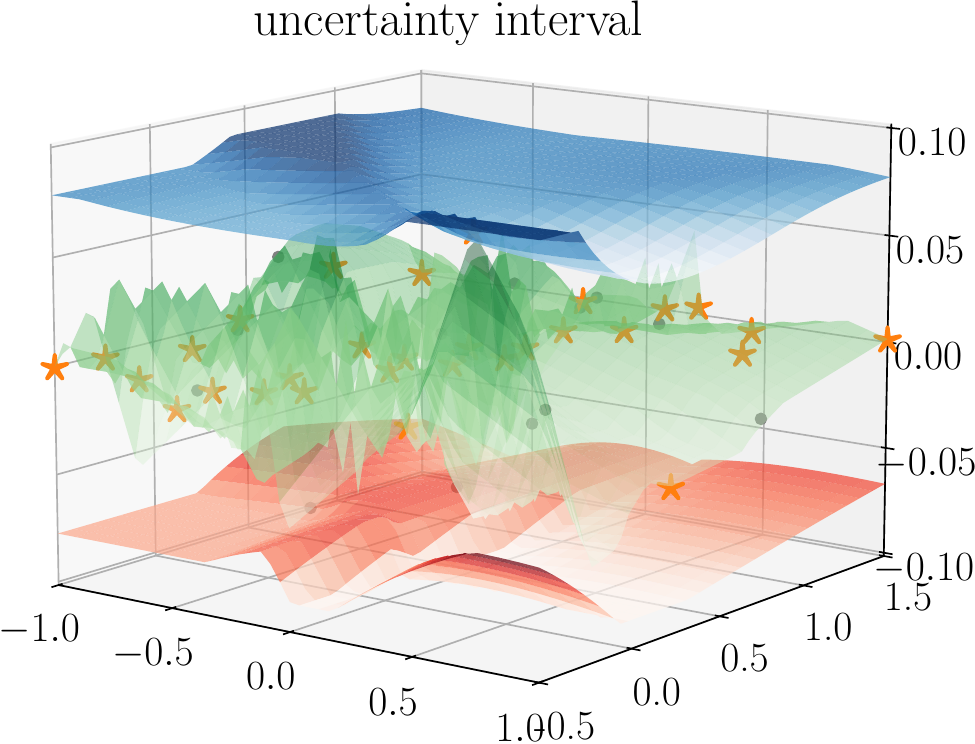}~\includegraphics[width=.48\hsize]{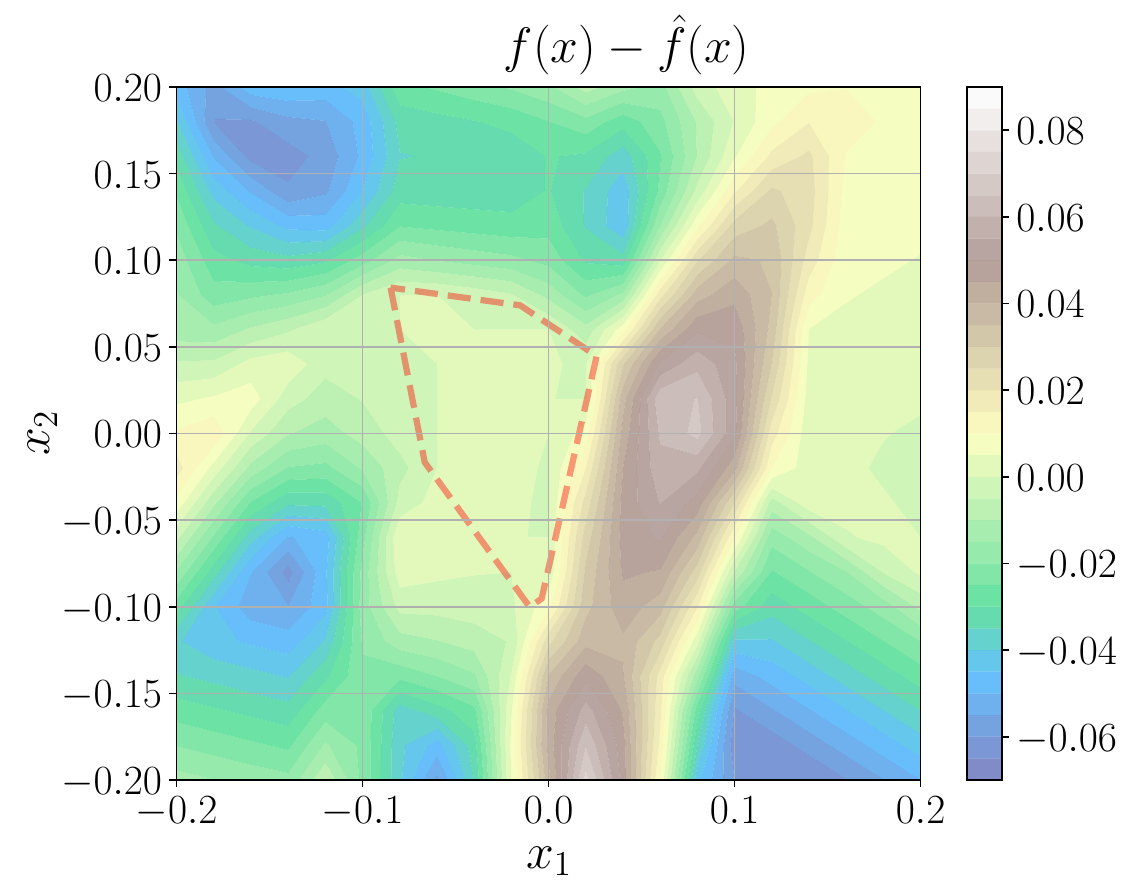}
\end{center}
\caption{mpQP example. Top-left: exact QP solution $f(x)$; Top-right: learned ReLU $\hat f(x;\theta)$; Bottom-left: pointwise error $e(x)=f(x)-\hat f(x;\theta)$ and envelope $[\bar e^\star_{\min}(x),\bar e^\star_{\max}(x)]$; Bottom-right: level sets of $f(x)-\hat f(x)$ around the origin and unconstrained-solution region~\eqref{eq:CR0} of mpQP problem (dashed-red polygon).}
\label{fig:mpqp-1}  
\end{figure}

We initialize Algorithm~\ref{algo:max_error_fit} with $N_0=20$ samples for $M=30$ active-learning iterations. Each sample is generated by solving the associated QP problem using the solver~\cite{ABA22b}. The total CPU time is $158.0$\,s ($147.2$\,s for training, $10.4$\,s for global optimization, $7.6$\,s to compute the asymmetric input-dependent confidence interval), with bound functions $\NN(x;\psi)$ also in $\FF$. The obtained results are shown in Figure~\ref{fig:mpqp-1}. Note that, as shown in the bottom-right plot of Figure~\ref{fig:mpqp-1}, the approximation error is close to zero in the region where no constraints of the mpQP are active, as expected. 

\subsection{Approximate linear MPC}
Consider the problem of controlling the non-minimum phase system with transfer function
$G(s)=(s-0.5)/(s^2+0.4s+1)$ from the input $u$ to the output $\tau$, under constraints
$-\dumax\leq \Delta u(t)\leq \dumax$, $\dumax=0.5$, 
and $-\tau_{\rm max}\leq \tau(t)\leq \tau_{\rm max}$, $\tau_{\rm max}=1.2$, for tracking setpoints $r(t)\in[-1,1]$. We consider the linear MPC formulation~\eqref{eq:linear-MPC} with prediction horizon $N=N_u=N_c=20$,
and weights $Q_\tau=1$, $Q_{\Delta u}=0.1$, $\rho_2=100$, $\rho_1=0$.
The resulting mpQP problem~\eqref{eq:mpQP} has $x\in\rr^4$ as parameter vector (2 states, 1 reference, 1 previous input), $z\in\rr^{21}$ as optimization vector (future control moves and slack), and 80 linear inequalities. 

\begin{figure}[t]
\begin{center}
\includegraphics[width=\hsize]{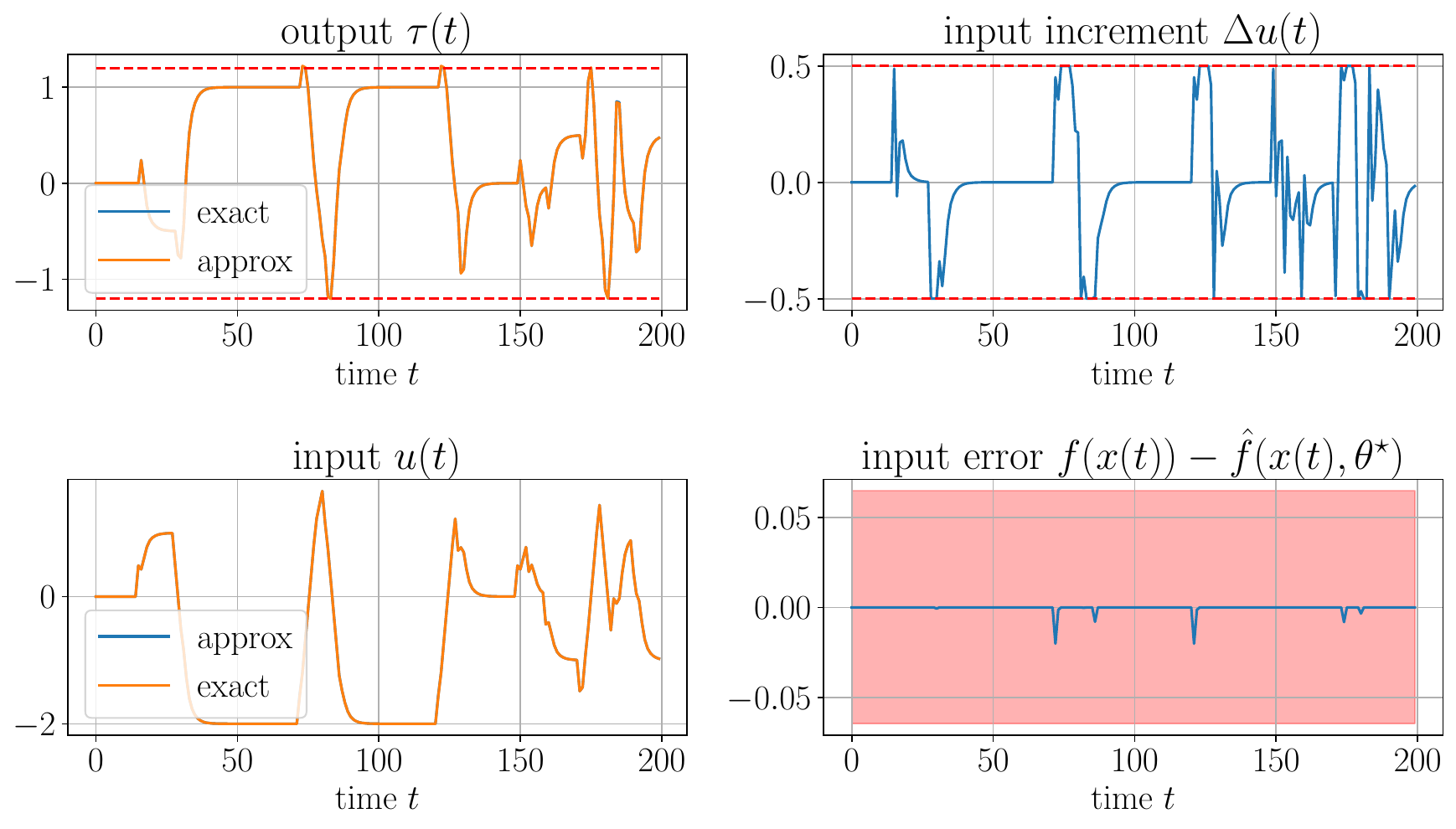}
\end{center}
\caption{Linear MPC example: closed-loop simulations under exact (QP-based) MPC control
and approximate MPC $\hat f(x,\theta^\star)$. The bottom-right plot shows the pointwise error $e(x)=f(x)-\hat f(x;\theta^\star)$ and symmetric uniform bound $[-\bar e^\star,\bar e^\star]$ returned by Algorithm~\ref{algo:max_error_fit}.}
\label{fig:mpc-1}
\end{figure}

We want to approximate the resulting MPC control law $y=f(x)$ using the approach of Section~\ref{sec:mpqp}
with a NN model $\NN(x;\theta)$ as in~\eqref{eq:mpqp-network}, except that we use two layers of 20 and 10 neurons, respectively, replace $\zmin=-\dumax+u_{-1}$, $\zmax=\dumax+u_{-1}$ in the saturation function to enforce input increment constraints, and collect samples to minimize the worst-case approximation error. To this end, we initialize Algorithm~\ref{algo:max_error_fit} with $N_0=1000$ samples generated by LHS in $\XX=\{x: [-3\ -3\ -1\ -2.5]^\top\leq x\leq [3\ 3\ 1\ 2.5]^\top\}$, which is an over-estimation of the region of interest for the system based on closed-loop MPC simulations for different reference excitations, for $M=1000$ active-learning iterations, with regularization $r(\theta)=10^{-4}\|\theta\|^2$ and $\nu=10^{-4}$ in~\eqref{eq:nu-weight}. Each sample is generated by solving the associated QP problem using the solver~\cite{ABA22b}. The results, obtained in $2233.3$\,s ($1947.7$\,s spent on training, $280.9$\,s in global optimization), are shown in Figure~\ref{fig:mpc-1}. 
The resulting WCE is $\bar e^\star=6.4602\cdot 10^{-2}$, while the MSE is $2.8812\cdot 10^{-4}$.

For comparison, we train the same model {\it without active learning} using $10^5$ randomly-generated samples, with both MSE loss and the loss~\eqref{eq:softmax}. In the first case, we obtain an MSE = $6.2401\cdot 10^{-4}$ but a considerably larger WCE $\bar e=0.3989$ over $\XX$, while with the loss~\eqref{eq:softmax} we get an MSE = $5.2893\cdot 10^{-4}$ and still a large WCE = $0.3982$. The models were trained in $69.81$\,s and $79.29$\,s, respectively, using the \texttt{jax-sysid} package~\cite{Bem25} from $10$ different random initializations. 
This highlights that ($i$) only judging the quality of the approximation based on MSE can be misleading, and ($ii$) the importance of an active-learning approach when minimizing the WCE.

\section{Conclusions}
We have proposed a general active-learning framework for training surrogate models with guaranteed, and possibly minimal, worst-case error bounds, applicable to a broad class of modeling and control tasks. The guaranteed properties of the error bounds critically rely on the global optimizer's ability to find the true global maximum of the approximation error; since global optimization scales poorly with the problem dimension, this may limit the applicability of the approach to small-scale problems, unless dedicated global optimizers are used for specific problems.

\appendix

\section{Proof of Proposition~\ref{prop:softmax}}
\label{app:softmax}
Consider any given vector $\theta\in\rr^{n_\theta}$, 
let $e_k(\theta)\eqdef y_k-\hat f(x_k;\theta)$, $\tilde e(\theta)\eqdef\max_k |e_k(\theta)|$, and let $\tilde K(\theta)$ be the set of indices $k$ such that $|e_k(\theta)|=\tilde e(\theta)$. Denote
by $\#\tilde K(\theta)$ the cardinality of $\tilde K(\theta)$ (the set $\tilde K(\theta)$ may contain more than one element, in general), and let $c\eqdef\#\tilde K$ if $\tilde e>0$ or $c\eqdef 2\#\tilde K$ if $\tilde e=0$.

By taking the limit $\gamma\to +\infty$ of the cost function in~\eqref{eq:softmax}, without regularization and dropping the dependence on $\theta$ for simplicity of notation, 
we get 
\[
    \begin{aligned}
    \lim_{\gamma\to +\infty}\! \frac{1}{\gamma}\log\left(\sum_{k=1}^{N} 
    e^{\gamma e_k}+e^{-\gamma e_k}\right)
    = \lim_{\gamma\to +\infty} \tilde e + \frac{1}{\gamma}\log\big(\sum_{k=1}^{N}
    \big(e^{\gamma(e_k-\tilde e)} + e^{\gamma (-e_k-\tilde e)}\big)\big)\\
    = \lim_{\gamma\to +\infty} \tilde e + \frac{1}{\gamma} \log\Big(c+\sum_{k\not\in \tilde K(\theta)}\big(e^{\gamma(e_k-\tilde e)} + e^{\gamma (-e_k-\tilde e)}\big)\Big)
    =\tilde e = \max_k|e_k|.
    \end{aligned}
\]
\cvd

\section{Proof of Proposition~\ref{prop:softsat}}
\label{app:softsat}
Consider the scalar case $y\in\rr$ for simplicity
and a generic $\eta>0$. Clearly, by setting $\bar y\eqdef \frac{(\ymax-\ymin)}{2}$, we get
$\softsat_\eta\left(\frac{\ymin+\ymax}{2};\ymin,\ymax\right)$ $=$
$\ymax+\frac{1}{\eta}\log\Big(\frac{1+e^{-\eta\bar y}}{1+e^{\eta\bar y}}\Big)$
$=$ $\ymax+\frac{1}{\eta}\log\left(e^{-\eta\bar y}\right)$ $=$ $\ymax-\bar y
    =\frac{(\ymax+\ymin)}{2}$.
The derivative of $\softsat_\eta(y;\ymin,\ymax)$ with respect to $y$ is

\[
    \softsat'_\eta(y;\ymin,\ymax)=\frac{e^{\eta (y-\ymin)}(1-e^{-\eta(\ymax-\ymin)})}{(1+e^{\eta(y-\ymin)})(1+e^{\eta(y-\ymax)})}
\]
whose denominator is positive. Since $\ymax>\ymin$ and $\eta>0$, its numerator is also positive
and, hence, $\softsat'_\eta(y;\ymin,\ymax)> 0$ for all $y\in\rr$, which implies that $\softsat_\eta(y;\ymin,\ymax)$ is monotonically strictly increasing
with respect to $y$.
Moreover, the limits of $\softsat_\eta(y;\ymin,\ymax)$ in~\eqref{eq:smooth_sat} are
\[
\begin{aligned}
&\lim_{y\to -\infty}\!\softsat_\eta(y;\ymin,\ymax) =\ymax + \frac{\log(e^{\eta(\ymin-\ymax)})}{\eta} = \ymin\\
&\lim_{y\to +\infty}\!\softsat_\eta(y;\ymin,\ymax) = \ymax+\frac{\log(1)}{\eta} = \ymax.
\end{aligned}
\]
If $\softsat_\eta(y;\ymin,\ymax)>\ymax$ or $\softsat_\eta(y;\ymin,\ymax)<\ymin$ were satisfied for some $y\in\rr$, then
$\softsat$ would not be monotonically increasing with respect to $y$, which is a contradiction.
Finally, for any $y\in(\ymin,\ymax)$, it is immediate to verify that $\lim_{\eta\rightarrow\infty} \softsat'_\eta(y;\ymin,\ymax)=1$.
\cvd

\end{document}